\PassOptionsToPackage{dvipsnames,table}{xcolor}   %
\documentclass[sigconf,10pt,nonacm]{acmart}
\setcopyright{none}
\renewcommand\footnotetextcopyrightpermission[1]{}

\usepackage{soul}         %
\usepackage{subcaption}
\usepackage[font=small]{caption}
\usepackage{multirow}
\usepackage{makecell}
\usepackage{enumitem}
\usepackage{tikz}
\usetikzlibrary{arrows.meta}
\usepackage{pifont}
\usepackage{array}
\usepackage{booktabs}
\usepackage{tabularx}
\usepackage{fvextra}     %

\definecolor{fsPurple}{HTML}{AF93C3}
\definecolor{fsBlue}{HTML}{84BCE1}
\definecolor{fsRed}{HTML}{E68A88}
\definecolor{fsLightPurple}{HTML}{E8DDEB}
\definecolor{fsLightBlue}{HTML}{C5DBE9}
\definecolor{fsLightRed}{HTML}{F4B7B7}
\definecolor{fsLightGreen}{HTML}{CFE5B9}
\definecolor{fsLightYellow}{HTML}{FBEFA6}
\definecolor{fsLightOrange}{HTML}{FAD7A8}
\definecolor{fsOrange}{HTML}{F6BE80}
\definecolor{fsGreen}{HTML}{80C080}

\colorlet{swfill}{fsLightBlue}              %
\colorlet{swline}{fsBlue!55!black}
\colorlet{convfill}{fsLightOrange}          %
\colorlet{netfill}{fsLightGreen}            %
\colorlet{hwline}{black!65}
\colorlet{txcol}{fsRed!80!black}            %
\colorlet{rxcol}{fsBlue!70!black}           %
\colorlet{ctlcol}{fsPurple!75!black}        %
\colorlet{cablecol}{fsOrange!85!black}      %

\begin{document}

\definecolor{lightgray}{gray}{0.9}
\newcommand\couldremove[1]{{\color{lightgray} #1}}
\newcommand {\remove}[1]{}

\newif\ifshowcomments
\showcommentstrue
\newcommand {\hlc}[2][yellow]{ {\sethlcolor{#1} \hl{#2}} }
\newcommand\note[1]{\ifshowcomments\hlc[SkyBlue]{-- #1 --}\fi}
\newcommand\tbd[1]{\colorbox{yellow}{\parbox{0.95\linewidth}{\small\textbf{TBD:} #1}}}
\newcommand\mynote[1]{\ifshowcomments\hlc[yellow]{#1}\fi}
\newcommand\tingjun[1]{\ifshowcomments\hlc[yellow]{TC: #1}\fi}
\newcommand\wei[1]{\ifshowcomments\hlc[pink]{WC: #1}\fi}
\newcommand\bill[1]{\ifshowcomments\hlc[cyan]{Bill: #1}\fi}

\newcommand\yiming[1]{\ifshowcomments\hlc[orange]{Yiming: #1}\fi}

\newcommand {\myparatight}[1]{\vspace{0.5ex}\noindent\textbf{#1~~}}
\newcommand {\specialcell}[2][c]{%
  \begin{tabular}[#1]{@{}c@{}}#2\end{tabular}}
\newcommand{\mycode}[1]{\texttt{\small #1}}
\newcommand{\cmark}{\textcolor{green!60!black}{\ding{51}}}
\newcommand{\xmark}{\textcolor{red!85!black}{\ding{55}}}

\newcommand{\projectcodetitle}{\texttt{uhd-rfsoc}}
\newcommand{\projectcode}{\mycode{uhd-rfsoc}}

\newcommand{\gsps}{GS/s}       %
\newcommand{\msps}{MS/s}
\newcommand{\gbps}{Gb/s}
\newcommand{\GbE}{GbE}

\newcommand{\Frfdc}{f_{\text{RFDC}}}
\newcommand{\Fsamp}{f_{s}}
\newcommand{\Fcarr}{f_{c}}
\newcommand{\bw}{B}
\newcommand{\Fint}{F_{\text{int}}}
\newcommand{\Fdec}{F_{\text{dec}}}
\newcommand{\Fintdec}{F}
\newcommand{\Rotw}{R_{\text{otw}}}

\DeclareRobustCommand{\ratebox}[2]{{\setlength{\fboxsep}{0.5pt}\colorbox{#1}{#2}}}
\DeclareRobustCommand{\rateA}[1]{#1}   %
\DeclareRobustCommand{\rateB}[1]{\ratebox{fsLightYellow}{#1}}
\DeclareRobustCommand{\rateC}[1]{\ratebox{fsLightOrange}{#1}}

\newcommand{\tabpre}{0mm}
\newcommand{\tabpost}{-4mm}
\newcommand{\figpost}{-4mm}

\newcommand*\circledwhite[1]{\tikz[baseline=(char.base)]{
            \node[shape=circle,draw,inner sep=0.6pt] (char) {\small{#1}};}}

\title{{\projectcodetitle}: Bringing Off-the-Shelf RFSoC Boards to Unmodified UHD Applications}

\author{Wei Cheng, Yiming Li, Junyao Zheng, Cheng-Bin Chen, and Tingjun Chen}
\affiliation{%
  \vspace{1mm}
  \institution{Department of Electrical and Computer Engineering, Duke University \vspace{2mm} \\
  {\url{https://github.com/functions-lab/uhd-rfsoc}}}
  \city{}
  \state{}
  \country{}
  }

\begin{abstract}
Next-generation wireless research relies on software-defined radios (SDRs) with more channels and larger bandwidth.
The USRP family is one of the most widely used SDR platforms, and much of the wireless software ecosystem is built on top of the USRP hardware driver (UHD).
However, each USRP model comes with a fixed channel count, bandwidth, and streaming capability.
Off-the-shelf radio frequency system-on-chip (RFSoC) boards, on the other hand, integrate up to 16 analog-to-digital converters (ADCs) and 16 digital-to-analog converters (DACs) operating at several gigasamples per second (GS/s), but existing UHD applications cannot directly use them.
We present {\projectcode}, an open-source FPGA design and host software stack that turns these boards, including the 16-channel ZCU216 and the RFSoC~4x2, into standard UHD devices without modifying UHD.
Through standard UHD calls, applications can configure the carrier frequency, channel count, the per-channel I/Q sample rate, and the wire format.
{\projectcode} combines timestamped, flow-controlled sample transport in the FPGA with data plane development kit (DPDK) packet processing and vectorized sample conversion on x86 and Arm hosts.
The built bitstreams for a number of RFSoC board configurations with varying RF data converter rates, channel counts, I/Q sample rates, and wire formats all successfully pass UHD's built-in streaming benchmarks.
We further demonstrate compatibility with unmodified GNU~Radio flowgraphs, a custom GPU-accelerated transceiver digital signal processing pipeline, as well as the OCUDU and Duranta 5G radio access network stacks using both software and commodity user equipment.
\end{abstract}

\maketitle

\section{Introduction}
\label{sec:intro}

\begin{figure*}[!t]
    \centering
    \includegraphics[width=0.93\textwidth]{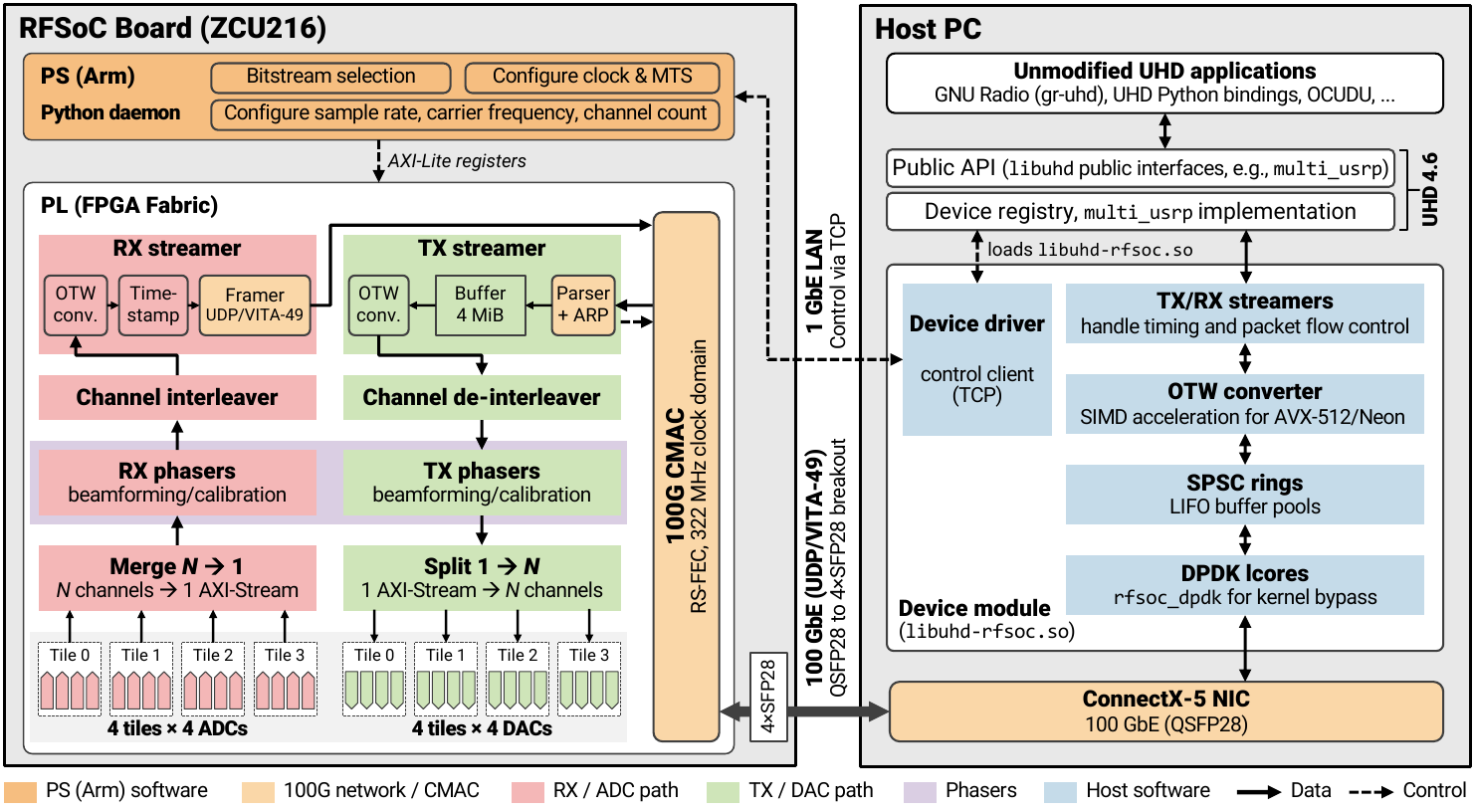}
    \vspace{-3.0mm}
    \caption{{\projectcode} at a glance.
    The RFSoC ZCU216 board (\emph{left}) and the host PC (\emph{right}) are connected via a 100\,GbE link carrying I/Q samples as UDP/VITA-49 packets and a 1\,GbE LAN carrying control traffic.
    In the FPGA, the RX path merges the $N$ selected ADC channels, applies the per-channel phasers, and interleaves and packetizes them; the TX path parses packets into a 4\,MiB buffer, reports its free space to the host, and applies the phasers before splitting it to the DAC channels.
    OTW Converters in the FPGA and on the host (with AVX-512 or Neon acceleration) translate samples to and from the selected wire format. 
    On the host, a DPDK-based device module (\texttt{libuhd-rfsoc.so}) serves unmodified UHD applications.
    The RFSoC~4x2 board (4 ADCs, 2 DACs, and one QSFP28) and the Arm host (DGX Spark, ConnectX-7) run the same design and module.}
    \label{fig:overview}
    \vspace{-4.0mm}
\end{figure*}

Next-generation (NextG) wireless networks are expected to support communication, sensing, and intelligence on a shared infrastructure, leveraging advanced radio techniques.
Examples include massive multiple-input, multiple-output (MIMO) systems~\cite{wang2024xlmimo} and communication in the upper-mid (frequency range 3, or FR~3) and millimeter-wave (mmWave) frequency bands~\cite{raychaudhuri2020challenge, kang2024upper, mezzavilla2024frequency, mezzavilla2025prototyping, chen2021programmable, gao2024mambas}, spectrum sensing and sharing~\cite{li2021enabling, subbaraman2023crescendo}, integrated sensing and communication (ISAC)~\cite{gao2026chameleon, pan2025nextg}, non-terrestrial networks~\cite{peng2025sigchord}, and AI-native radio access networks (AI-RAN)~\cite{yao2026beaver}.
All of these require high-performance radios: digital-to-analog and analog-to-digital converters (DACs and ADCs) and radio frequency (RF) front ends that bridge the RF world with the hardware platform performing digital signal processing (DSP).
At the same time, more of this processing is moving into software on commodity (edge) servers.
For example, virtualized RANs (vRANs) now meet the tight slot deadlines of massive MIMO and mmWave baseband processing using a pool of CPU cores, GPUs, and ASIC accelerators~\cite{ding2020agora, qi2024savannah, xie2025megastation, qi2026nexus}, and scale it to multiple cells on heterogeneous compute~\cite{schiavo2024cloudric, qi2026nexus}.

While the processing software now scales with commodity compute, the radio hardware does not scale as easily. 
NextG radio systems grow along two dimensions: more antennas to support massive MIMO and beamforming, and larger bandwidth, from hundreds of MHz to a few GHz, which requires streaming and processing hundreds of millions to billions of complex-valued I/Q samples per second.
Software-defined radios (SDRs) must therefore scale in both dimensions, while streaming samples reliably at high rates and flexibly partitioning DSP stages between the FPGA and the host.

Among existing SDR platforms, the USRP family is one of the most widely used in wireless research.
All USRP models can be accessed through the USRP Hardware Driver (UHD)~\cite{ettus_uhd}, an open-source driver and application programming interface (API) that allows applications to configure the SDR and to stream samples to/from it.
Since GNU Radio~\cite{gnuradio_gruhd}, OCUDU~\cite{ocudu}, Duranta~\cite{duranta}, and many other applications are built on top of UHD, it has become the \emph{de facto} software interface for SDRs: any radio that implements this interface immediately inherits the entire ecosystem.
However, each USRP model comes with a different channel count, bandwidth, and streaming capability, so scaling in either dimension requires several synchronized units or a different model.

Off-the-shelf RFSoC boards offer another path to scale.
AMD's RF System-on-Chip (RFSoC) integrates high-speed RF-ADCs and RF-DACs, a field-programmable gate array (FPGA), and Arm CPU cores on one chip. 
The board configurations range from 4 ADCs and 2 DACs on the RFSoC~4x2 kit~\cite{realdigital_rfsoc4x2} to 16 of each on the ZCU216 evaluation board~\cite{amd_zcu216}, with every converter sampling at several giga-samples per second ({\gsps}). 
They have enabled wideband radio testbeds~\cite{cheng2024spear, cheng2026spearbf, haque2024m3mimo, lacruz2021mimorph, ruiz2025helix} and high-rate sample streamers~\cite{engelhardt2024accelerating, siauciulis2023100gbit, heidari2026radiostream}.
However, each of these platforms comes with its own device driver and software stack, so stock UHD applications (e.g., the open-source RAN software stacks OCUDU~\cite{ocudu} and Duranta~\cite{duranta}) cannot use them (see Table~\ref{tab:sdrs}).

\begin{table*}[!t]
    \centering
    \caption{Comparison of existing USRP devices supported by UHD, RFSoC-based SDR platforms with customized drivers, and \texttt{uhd-rfsoc}.
    Each device includes its FPGA part, the number of TX and RX channels, the maximum I/Q sample rate per channel, the aggregated I/Q sample rate across all channels in one direction, the host-device link, and the supported over-the-wire (OTW) formats.
    \texttt{sc16}, \texttt{sc12}, and \texttt{sc8} denote 16-, 12-, and 8-bit integer I and Q samples, respectively.}
    \vspace{-3.0mm}
    \label{tab:sdrs}
    \begingroup
    \scriptsize
    \setlength{\tabcolsep}{4pt}
    \renewcommand{\arraystretch}{1.25}
    \begin{tabularx}{\linewidth}{@{}
        >{\hsize=0.60\hsize\raggedright\arraybackslash}X
        >{\hsize=0.65\hsize\raggedright\arraybackslash}X
        >{\hsize=0.85\hsize\raggedright\arraybackslash}X
        >{\hsize=0.60\hsize\raggedright\arraybackslash}X
        >{\hsize=1.00\hsize\raggedright\arraybackslash}X
        >{\hsize=0.75\hsize\centering\arraybackslash}X
        >{\hsize=0.80\hsize\centering\arraybackslash}X
        >{\hsize=0.60\hsize\centering\arraybackslash}X
        @{}}
        \toprule
        \textbf{Device driver}
        & \textbf{Device}
        & \textbf{FPGA part}
        & \textbf{\# of channels}
        & \textbf{Max sample rate per ch.}
        & \textbf{Agg. sample rate}
        & \textbf{Host-device link}
        & \textbf{OTW format} \\
        \midrule
        \texttt{uhd} & USRP X310~\cite{ettus_x310}
            & XC7K410T
            & 2 TX, 2 RX & 200\,{\msps} & 400\,{\msps} & 2$\times$10\,GbE & sc16$^{1}$ \\ [1pt]
         & USRP N310~\cite{ettus_n310}
            & XC7Z100
            & 4 TX, 4 RX & 100\,{\msps} & 400\,{\msps} & 2$\times$10\,GbE & sc16$^{1}$ \\ [1pt]
         & USRP X410~\cite{ettus_x410}
            & XCZU28DR (Gen 1)
            & 4 TX, 4 RX & 500\,{\msps} & 2.0\,{\gsps} & 2$\times$100\,GbE & sc16$^{1}$ \\ [1pt]
         & USRP X440~\cite{ettus_x440}
            & XCZU28DR (Gen 1)
            & 8 TX, 8 RX & \makecell[l]{500\,{\msps} (8 ch.), \\ 2.0\,{\gsps} (2 ch.)} & 4.0\,{\gsps} & 2$\times$100\,GbE & sc16$^{1}$ \\
        \midrule
        Custom driver$^{2}$ & RFSoC~4x2~\cite{siauciulis2023100gbit}
            & XCZU48DR (Gen 3)
            & 1 RX & 2.46\,{\gsps} & 2.46\,{\gsps} & 1$\times$100\,GbE & sc16 \\ [1pt]
         & Custom SDR~\cite{engelhardt2024accelerating}
            & XCZU48DR (Gen 3)
            & 1 TX, 1 RX & 2.5\,{\gsps} & 2.5\,{\gsps} & 1$\times$100\,GbE & sc16 \\ [1pt]
         & \makecell[l]{ZCU111, \\ ZCU208~\cite{ruiz2025helix}}
            & \makecell[l]{XCZU28DR (Gen 1), \\ XCZU48DR (Gen 3)}
            & 1 TX, 1 RX & 491.52\,{\msps} & 491.52\,{\msps} & 1$\times$10\,GbE & sc16 \\ [1pt]
         & 8T8R SOM~\cite{heidari2026radiostream}
            & XCZU48DR (Gen 3)
            & 8 TX, 8 RX & \makecell[l]{245.76\,{\msps} (8 ch.), \\ 491.52\,{\msps} (4 ch.)} & 1.97\,{\gsps} & 1$\times$100\,GbE & sc16 \\
        \midrule
        \textbf{\texttt{uhd-rfsoc}}
        & \textbf{ZCU216}
        & \textbf{XCZU49DR (Gen~3)}
        & \textbf{16 TX, 16 RX} & \makecell[l]{\textbf{156.25\,{\msps} (16 ch.)}, \\ \textbf{1.25\,{\gsps} (2 ch.)}} & \textbf{2.5\,{\gsps}}$^{3}$ & \textbf{1$\times$100\,GbE} & \textbf{sc16}/\textbf{sc12}/\textbf{sc8} \\ [5pt]
        & \textbf{RFSoC~4x2}
        & \textbf{XCZU48DR (Gen~3)}
        & \textbf{2 TX, 2 RX} & \makecell[l]{\textbf{1.25\,{\gsps} (2 ch.),} \\ \textbf{2.5\,{\gsps} (1 ch.)}} & \textbf{2.5\,{\gsps}}$^{3}$ & \textbf{1$\times$100\,GbE} & \textbf{sc16}/\textbf{sc12}/\textbf{sc8} \\
        \bottomrule
    \end{tabularx}
    \\ [3pt]
    \parbox{\linewidth}{\scriptsize
    $^{1}$\texttt{sc16} is the only wire format that UHD supports on USRP X3xx, N3xx, and X4xx devices~\cite{ettus_uhd_otw}.
    $^{2}$Board-specific host driver supporting GNU~Radio/Python/MATLAB without UHD support.
    $^{3}$Under \texttt{sc16} wire format.}
    \endgroup
    \vspace{-4.0mm}
\end{table*}

Bringing an off-the-shelf RFSoC board to UHD and supporting unmodified UHD applications is nontrivial.
First, the 100\,GbE link may carry only a fraction of the samples that the data converters can produce, and the system needs to let applications select a channel count, I/Q sample rate, and wire format that fit the link.
Second, the FPGA has to timestamp, buffer, and flow-control the samples for DACs and ADCs that run at a fixed rate.
Third, the host has to move and convert samples at tens of Gb/s in each direction, which general-purpose host software cannot easily do.
Moreover, all of this has to work without modifying UHD, so that existing applications run unchanged.

To address these challenges, we present {\projectcode}, shown in Fig.~\ref{fig:overview}, an FPGA design and host software stack that makes off-the-shelf RFSoC boards, such as the 16-channel ZCU216 and the RFSoC~4x2, compatible with UHD, allowing existing applications to access the RFSoC boards through the standard, unmodified UHD interfaces and APIs.
{\projectcode} does not modify UHD; its host device module is loaded by UHD as a new device type, and a daemon on the RFSoC's Arm core configures the data converters and the FPGA.
The system combines timestamped sample transport and flow-controlled buffering in the FPGA with data plane development kit (DPDK)~\cite{dpdk} packet processing and vectorized sample conversion on x86 and Arm hosts.
Through standard UHD calls, applications configure the carrier frequency, channel count, the per-channel I/Q sample rate, and the over-the-wire (OTW) format.
Over one 100\,GbE host-device link, {\projectcode} sustains an aggregate I/Q sample rate of 2.5\,{\gsps} (5.0\,{\gsps}) per direction with 16-bit (8-bit) I/Q samples, corresponding to a per-channel I/Q sample rate of 156.25\,{\msps} (312.5\,{\msps}) on the ZCU216 board with 16 channels, or 1.25\,{\gsps} (2.5\,{\gsps}) on the RFSoC~4x2 board with 2 channels.

We evaluate {\projectcode} using UHD's streaming benchmarks and demonstrate operation with unmodified GNU~Radio blocks and customized transceiver DSP pipelines, as well as the open-source OCUDU and Duranta 5G RAN stacks with both software-implemented and commodity user equipment (UE) across both x86 and Arm hosts.
The results show that with {\projectcode}, off-the-shelf RFSoC boards can support existing wireless software while extending the hardware configurations available to researchers.

In summary, this paper makes the following contributions:
\begin{itemize}[leftmargin=*, topsep=2pt, itemsep=1pt]
    \item
    We design an FPGA data plane and a board control daemon for off-the-shelf RFSoC boards that stream timestamped I/Q samples over one 100\,GbE host-device link as UDP/VITA-49 packets, with flow-controlled TX buffering and per-channel phasers for beamforming and calibration.
    Users select the channel count, per-channel I/Q sample rate, and wire formats (16-, 12-, or 8-bit) that trade sample resolution for bandwidth;
    \item
    We develop a host-side UHD device module that exposes the RFSoC boards to \emph{unmodified} UHD applications as a new device type.
    It accesses the network card directly through DPDK and converts samples between the wire format and the application's format with SIMD CPU instructions, AVX-512 on x86 and Neon on Arm hosts, while keeping up with the 100\,GbE host-device link;
    \item
    We comprehensively evaluate {\projectcode} on the ZCU216 and RFSoC~4x2 boards with unmodified UHD software and applications across four levels, from UHD's built-in benchmarks and GNU~Radio workflows to customized transceiver DSP pipelines and open-source RAN stacks.
\end{itemize}
\noindent
The {\projectcode} project, along with the compiled bitstreams for ZCU216 and RFSoC~4x2 boards, is open-sourced at~\cite{uhd_rfsoc_repo}. 

\section{Related Work}
\label{sec:related}

\myparatight{SDR platforms and the UHD ecosystem.}
The USRP family is one of the most widely used SDR platforms, and UHD~\cite{ettus_uhd} is its \emph{de facto} driver and API.
Many-antenna and massive MIMO radio testbeds~\cite{doost2018renew, shepard2012argos, ding2020agora} build large arrays from many synchronized SDR units.
The RFSoC-based USRP X410~\cite{ettus_x410} and X440~\cite{ettus_x440} natively provide UHD streaming, with DPDK transport across dual\,100 GbE links.
However, their channel counts and sample rates are set by the vendor's FPGA images: four channels of 500\,{\msps} on the X410, and eight channels of 500\,{\msps} or two channels of 2.0\,{\gsps} on the X440, of which the 100\,GbE images stream four full-duplex channels at 500\,{\msps} or two at 1.0\,{\gsps}.
RF Network-on-Chip, or RFNoC~\cite{braun2016rfnoc}, allows users to add DSP blocks to the FPGA within a USRP, but it targets USRP hardware and its CHDR packet format.
Hardware abstraction layers such as SoapySDR~\cite{soapysdr}, used by the Iris radios in RENEW~\cite{doost2018renew}, provide a common API across these drivers, and SoapyUHD~\cite{soapyuhd} can expose a SoapySDR device to UHD applications.

\myparatight{RFSoC-based platforms.}
AMD RFSoC platforms have emerged as a compelling foundation for wideband, mmWave, and massive MIMO experimental testbeds~\cite{dhananjay2021calibration, lacruz2021mimorph, haque2024m3mimo, sahin2023millimeter, cheng2024spear, cheng2026spearbf}, real-time 6G prototyping ~\cite{ruiz2025helix, engelhardt2024accelerating}, large-scale antenna arrays such as a 256-chain system composed of 16 synchronized ZCU216 boards~\cite{iancu2026scalable}, and quantum instrumentation controllers~\cite{stefanazzi2022qick}.
Existing platforms generally fall into two categories.
The first category streams I/Q samples continuously between the RFSoC data converters and the host over high-speed network interfaces.
For instance,~\cite{siauciulis2023100gbit} offloads one receiver channel at 2.46\,{\gsps} (about 80\,{\gbps}) into GNU Radio over kernel UDP sockets, but is restricted to receive-only streaming and requires customized GNU Radio blocks. 
RadioStream~\cite{heidari2026radiostream} streams up to eight channels at 245.76\,{\msps} or four channels at 491.52\,{\msps} (1.97\,{\gsps} in aggregate), with the transfer direction set per 0.5\,ms slot, over raw Ethernet frames with a DPDK controller, but relies on custom out-of-tree GNU Radio blocks.
The platform in~\cite{engelhardt2024accelerating} achieves 80\,{Gb/s} streaming using a C++/DPDK driver with a Python application layer, and HELIX~\cite{ruiz2025helix} provides real-time PHY processing over a customized software stack.
The second category avoids continuous streaming by operating on bounded sample buffers, using either MATLAB over TCP~\cite{dhananjay2021calibration, lacruz2021mimorph} or Python on the board's embedded Arm subsystem via PYNQ~\cite{goldsmith2020control, sahin2023millimeter, cheng2024spear}, or keep the DSP entirely in the FPGA and use PYNQ only for control and decimated monitoring~\cite{iancu2026scalable}.
In these systems, waveforms are pre-loaded into or captured from on-chip block RAM (BRAM) or external DRAM for offline analysis.
While effective for static channel sounding, array calibration, and packet-level experiments, this burst-oriented approach cannot sustain continuous, real-time full-duplex traffic, which prevents the evaluation of closed-loop MAC/PHY protocols or live cellular RAN stacks.
In both categories, each platform comes with its own host software, so existing UHD applications cannot run on it.

\myparatight{High-rate sample transport on the host.}
At tens of Gb/s, the host-side data path becomes the bottleneck, and kernel UDP sockets cannot easily keep up with a 100\,GbE link.
UHD already supports DPDK~\cite{dpdk} for high-rate USRP transport and provides a public, extensible sample conversion interface.
RDMA over Converged Ethernet (RoCE) has been proposed to eliminate host packet-processing overhead entirely, though experimental evaluations remain open~\cite{andrich2026beyond}.

\section{Design of {\projectcodetitle}}
\label{sec:design}

\begin{table}[t]
\centering
\caption{Notation and configuration parameters of {\projectcode}.}
\label{tab:terms}
\footnotesize
\setlength{\tabcolsep}{6pt}
\renewcommand{\arraystretch}{1.25}
\vspace{-3.0mm}
\begin{tabularx}{\linewidth}{@{}
        >{\hsize=0.14\hsize\raggedleft\arraybackslash}X
        >{\hsize=0.80\hsize\raggedright\arraybackslash}X
        @{}}
\toprule
\textbf{Symbol} & \textbf{Definition} \\
\midrule
$\Frfdc$ & Sampling rate of the RF data converter (RFDC) \\
$\Fintdec$ & Data converter interpolation (TX) or decimation (RX) factor \\
$\Fsamp$ & I/Q sample rate per channel, $\Fsamp = \Frfdc / \Fintdec$ \\
$\Fcarr$ & Carrier frequency set by the RF data converter numerically controlled oscillator (NCO), $0 \le \Fcarr \le \Frfdc/2$ \\
$N$ & Number of streamed channels per direction \\
$b$ & Bits per I/Q sample streamed over the wire: 32 (\texttt{sc16}), 24 (\texttt{sc12}), or 16 (\texttt{sc8}) \\
$\Rotw$ & Aggregated I/Q sample payload rate, $\Rotw = N \cdot \Fsamp \cdot b$ \\
\bottomrule
\end{tabularx}
\vspace{-5.0mm}
\end{table}

\begin{table*}[!t]
    \centering
    \caption{Example configurations supported by {\projectcode} on the ZCU216 and RFSoC~4x2 devices, including the RF data converter (RFDC) sample rates $\Frfdc$, carrier frequency range $\Fcarr$ (without external RF front ends), channel count $N$, and per-channel I/Q sample rates.
    The supported sample rates $\Fsamp$ are derived from $\Frfdc$ using the integer interpolation (TX) or decimation (RX) factor $\Fintdec$; here $\Fsamp = \Frfdc/\Fintdec$ plays the role of the master clock rate (MCR) of a USRP device.
    For each wire format (\mycode{sc16}, \mycode{sc12}, and \mycode{sc8}), the maximum sample rate is the highest per-channel I/Q sample rate whose aggregate stream across all $N$ channels fits the 100\,GbE host-device link.
    The 2.4576\,{\gsps} and 4.9152\,{\gsps} RFDC clocks provide LTE/5G~NR sample rates (bold), and further I/Q sample rates could be supported by adding fabric digital up-/down-conversion (DUC/DDC).
    Each row corresponds to one compiled bitstream.
    In the last column, $\Fsamp$ that fit the 100\,GbE link across all $N$ channels only with a narrower wire format are highlighted: \rateB{yellow} for \mycode{sc12} or \mycode{sc8}, and \rateC{orange} for \mycode{sc8} only; the others fit in any wire format, including \mycode{sc16}.}
    \label{tab:rfsoc-board-config}
    \vspace{\tabpre}
    \vspace{-3.0mm}
    \scriptsize
    \setlength{\tabcolsep}{3pt}
    \begin{tabularx}{\linewidth}{@{}
        >{\hsize=0.10\hsize\raggedright\arraybackslash}X
        >{\hsize=0.11\hsize\centering\arraybackslash}X
        >{\hsize=0.14\hsize\centering\arraybackslash}X
        >{\hsize=0.06\hsize\centering\arraybackslash}X
        >{\hsize=0.22\hsize\raggedright\arraybackslash}X
        >{\hsize=0.70\hsize\raggedright\arraybackslash}X
        @{}}
        \toprule
        \textbf{Device} & \makecell[c]{\textbf{RFDC rate} \\ \textbf{($\Frfdc$)}} & \makecell[c]{\textbf{Freq. coverage} \\ ($\Fcarr$)} & $N$ & \makecell[l]{\textbf{Max sample rate per ch.} \\ \textbf{in \texttt{sc16}, \texttt{sc12}, \texttt{sc8} ({\msps})}} & \textbf{Supported sample rates per ch. ($\Fsamp$, {\msps})} \\
        \midrule
        \textbf{ZCU216} & 2.5\,GS/s & {0--1250}\,MHz & 1 & 1250, 1250, 1250 & \rateA{62.5}, \rateA{104.17}, \rateA{125}, \rateA{156.25}, \rateA{208.33}, \rateA{250}, \rateA{312.5}, \rateA{416.67}, \rateA{500}, \rateA{625}, \rateA{833.33}, \rateA{1250} \\
        & & & 2 & 1250, 1250, 1250 & \rateA{62.5}, \rateA{104.17}, \rateA{125}, \rateA{156.25}, \rateA{208.33}, \rateA{250}, \rateA{312.5}, \rateA{416.67}, \rateA{500}, \rateA{625}, \rateA{833.33}, \rateA{1250} \\
        &  &  & 4 & 625, 833.33, 1250 & \rateA{62.5}, \rateA{104.17}, \rateA{125}, \rateA{156.25}, \rateA{208.33}, \rateA{250}, \rateA{312.5}, \rateA{416.67}, \rateA{500}, \rateA{625}, \rateB{833.33}, \rateC{1250} \\
        &  &  & 8 & 312.5, 500, 625 & \rateA{62.5}, \rateA{104.17}, \rateA{125}, \rateA{156.25}, \rateA{208.33}, \rateA{250}, \rateA{312.5}, \rateB{416.67}, \rateB{500}, \rateC{625} \\
        &  &  & 16 & 156.25, 250, 312.5 & \rateA{62.5}, \rateA{104.17}, \rateA{125}, \rateA{156.25}, \rateB{208.33}, \rateB{250}, \rateC{312.5} \\
        \cmidrule(lr){2-6}
        & 2.4576\,{\gsps} & {0--1228.8}\,MHz & 1 & 1228.8, 1228.8, 1228.8 & \rateA{\textbf{61.44}}, \rateA{102.4}, \rateA{\textbf{122.88}}, \rateA{153.6}, \rateA{204.8}, \rateA{\textbf{245.76}}, \rateA{307.2}, \rateA{409.6}, \rateA{\textbf{491.52}}, \rateA{614.4}, \rateA{819.2}, \rateA{1228.8} \\
        & & & 2 & 1228.8, 1228.8, 1228.8 & \rateA{\textbf{61.44}}, \rateA{102.4}, \rateA{\textbf{122.88}}, \rateA{153.6}, \rateA{204.8}, \rateA{\textbf{245.76}}, \rateA{307.2}, \rateA{409.6}, \rateA{\textbf{491.52}}, \rateA{614.4}, \rateA{819.2}, \rateA{1228.8} \\
        &  &  & 4 & 614.4, 819.2, 1228.8 & \rateA{\textbf{61.44}}, \rateA{102.4}, \rateA{\textbf{122.88}}, \rateA{153.6}, \rateA{204.8}, \rateA{\textbf{245.76}}, \rateA{307.2}, \rateA{409.6}, \rateA{\textbf{491.52}}, \rateA{614.4}, \rateB{819.2}, \rateC{1228.8} \\
        &  &  & 8 & 307.2, 491.52, 614.4 & \rateA{\textbf{61.44}}, \rateA{102.4}, \rateA{\textbf{122.88}}, \rateA{153.6}, \rateA{204.8}, \rateA{\textbf{245.76}}, \rateA{307.2}, \rateB{409.6}, \rateB{\textbf{491.52}}, \rateC{614.4} \\
        &  &  & 16 & 153.6, 245.76, 307.2 & \rateA{\textbf{61.44}}, \rateA{102.4}, \rateA{\textbf{122.88}}, \rateA{153.6}, \rateB{204.8}, \rateB{\textbf{245.76}}, \rateC{307.2} \\
        \cmidrule(lr){2-6}
        & 2.4\,{\gsps} & {0--1200}\,MHz & 1 & 1200, 1200, 1200 & \rateA{60}, \rateA{100}, \rateA{120}, \rateA{150}, \rateA{200}, \rateA{240}, \rateA{300}, \rateA{400}, \rateA{480}, \rateA{600}, \rateA{800}, \rateA{1200} \\
        & & & 2 & 1200, 1200, 1200 & \rateA{60}, \rateA{100}, \rateA{120}, \rateA{150}, \rateA{200}, \rateA{240}, \rateA{300}, \rateA{400}, \rateA{480}, \rateA{600}, \rateA{800}, \rateA{1200} \\
        &  &  & 4 & 600, 800, 1200 & \rateA{60}, \rateA{100}, \rateA{120}, \rateA{150}, \rateA{200}, \rateA{240}, \rateA{300}, \rateA{400}, \rateA{480}, \rateA{600}, \rateB{800}, \rateC{1200} \\
        &  &  & 8 & 300, 480, 600 & \rateA{60}, \rateA{100}, \rateA{120}, \rateA{150}, \rateA{200}, \rateA{240}, \rateA{300}, \rateB{400}, \rateB{480}, \rateC{600} \\
        &  &  & 16 & 150, 240, 300 & \rateA{60}, \rateA{100}, \rateA{120}, \rateA{150}, \rateB{200}, \rateB{240}, \rateC{300} \\
        \midrule
        \textbf{RFSoC~4x2} & 5.0\,GS/s & {0--2500}\,MHz & 1 & 2500, 2500, 2500 & \rateA{125}, \rateA{208.33}, \rateA{250}, \rateA{312.5}, \rateA{416.67}, \rateA{500}, \rateA{625}, \rateA{833.33}, \rateA{1000}, \rateA{1250}, \rateA{1666.67}, \rateA{2500} \\
        &  &  & 2 & 1250, 1666.67, 2500 & \rateA{125}, \rateA{208.33}, \rateA{250}, \rateA{312.5}, \rateA{416.67}, \rateA{500}, \rateA{625}, \rateA{833.33}, \rateA{1000}, \rateA{1250}, \rateB{1666.67}, \rateC{2500} \\
        \cmidrule(lr){2-6}
        & 4.9152\,GS/s & {0--2457.6}\,MHz & 1 & 2457.6, 2457.6, 2457.6 & \rateA{\textbf{122.88}}, \rateA{204.8}, \rateA{\textbf{245.76}}, \rateA{307.2}, \rateA{409.6}, \rateA{\textbf{491.52}}, \rateA{614.4}, \rateA{819.2}, \rateA{983.04}, \rateA{1228.8}, \rateA{1638.4}, \rateA{2457.6} \\
        &  &  & 2 & 1228.8, 1638.4, 2457.6 & \rateA{\textbf{122.88}}, \rateA{204.8}, \rateA{\textbf{245.76}}, \rateA{307.2}, \rateA{409.6}, \rateA{\textbf{491.52}}, \rateA{614.4}, \rateA{819.2}, \rateA{983.04}, \rateA{1228.8}, \rateB{1638.4}, \rateC{2457.6} \\
        \cmidrule(lr){2-6}
        & 4.8\,{\gsps} & {0--2400}\,MHz & 1 & 2400, 2400, 2400 & \rateA{120}, \rateA{200}, \rateA{240}, \rateA{300}, \rateA{400}, \rateA{480}, \rateA{600}, \rateA{800}, \rateA{960}, \rateA{1200}, \rateA{1600}, \rateA{2400} \\
        &  &  & 2 & 1200, 1600, 2400 & \rateA{120}, \rateA{200}, \rateA{240}, \rateA{300}, \rateA{400}, \rateA{480}, \rateA{600}, \rateA{800}, \rateA{960}, \rateA{1200}, \rateB{1600}, \rateC{2400} \\
        \bottomrule
    \end{tabularx}
    \vspace{\tabpost}
\end{table*}

{\projectcode} connects off-the-shelf RFSoC boards to applications through the standard UHD interfaces and APIs~\cite{ettus_uhd}.
Exposing an RFSoC board as a UHD device means supporting various device configurations and sustaining reliable, high-rate I/Q streaming while keeping UHD itself unmodified.
This raises challenges on both the FPGA and host sides.

\myparatight{Challenges on the FPGA side.}
First, the host-device link, instead of the data converters, limits the streamed I/Q sample rate.
At $\Frfdc = 2.5\,\text{GS/s}$, the 16 ADCs of the ZCU216 together produce up to 20\,{\gsps} of I/Q samples, or 640\,{Gb/s} in \mycode{sc16}, which is more than six times the capacity of the 100 GbE link.
The FPGA design therefore needs to support different channel counts, I/Q sample rates, and wire formats, so that an application can select a configuration that fits the link (\S\ref{sec:design:config}).
Second, the DACs and ADCs run at a fixed rate and cannot wait for the host.
On the TX side, the 4\,{MiB} on-chip buffer holds only about 0.4\,ms of samples at 80\,Gb/s, so the FPGA has to pace the host to avoid both overrunning and draining the buffer.
On the RX side, the FPGA cannot apply backpressure to the ADCs, so any sample loss has to be reported to the host rather than silently corrupting the stream.
Third, UHD applications expect timestamped samples, timed stream commands, and time-aligned channels.
The FPGA fabric and the board daemon have to provide these across multiple RFDC tiles, and keep them when the application changes the I/Q sample rate (\S\ref{sec:design:fpga} and \S\ref{sec:design:daemon}).

\myparatight{Challenges on the host side.}
At 80\,Gb/s per direction, a packet of 2,048 I/Q samples arrives every 0.82\,µs, and general-purpose host software cannot keep up with this rate.
First, the kernel network stack and a scalar implementation of the OTW converter both run well below the line rate.
Second, memory bandwidth runs out before computation. Full-duplex streaming moves giga-bytes of I/Q samples per second through host memory, and DPDK's default buffer pools recycle buffers in first-in, first-out (FIFO) order, so a newly allocated buffer is the one least likely to still be in the cache.
Third, the operating system schedules its own workers, periodic jobs, and interrupts on every core, and each preemption of a busy-polling core is long enough to stall the sample stream.
If such a stall outlasts the FPGA's TX buffer (e.g., while a worker thread sleeps until a timed burst starts), the buffer runs dry and the DACs underrun.
{\projectcode} addresses these challenges in its host device module with DPDK packet processing, vectorized OTW conversion, LIFO buffer recycling, and reserved streaming cores (\S\ref{sec:design:host}).

\myparatight{Compatibility with unmodified UHD.}
UHD's own transport, streamers, and sample converters are private to \mycode{libuhd} and tied to its CHDR packet format, so a device with a different wire format cannot reuse them.
{\projectcode} therefore enters UHD only through its device registry and provides these layers in a separate device module (\S\ref{sec:design:host}), while a daemon on the board owns all FPGA registers (\S\ref{sec:design:daemon}).

The architecture of {\projectcode} consists of three components, as shown in Fig.~\ref{fig:overview}:
an FPGA data plane in the RFSoC's programmable logic (PL), a control daemon running on the board's Arm processor, and a UHD device module tailored for RFSoC devices running on the host, described next.

\subsection{RFSoC Device Configuration}
\label{sec:design:config}

Three parameters define the configuration of an RFSoC device: the channel count $N$, the per-channel I/Q sample rate $\Fsamp$, and the wire format. 
For example, the ZCU216 board supports up to 16 concurrent TX and RX channels; the RFSoC~4x2 board has two DACs and four ADCs, of which {\projectcode} streams up to two TX and two RX channels.
An I/Q sample rate of 1\,MS/s corresponds to 1\,MHz of nominal bandwidth, and we use MS/s (or GS/s) when specifying I/Q sample rates and MHz (or GHz) when specifying channel bandwidth.

\myparatight{I/Q sample rate and host-device link budget.}
For a given sample rate of the RF data converter (RFDC), $\Frfdc$, the converters decimate on RX or interpolate on TX by an integer factor $\Fintdec$, yielding an I/Q sample rate of $\Fsamp = \Frfdc/\Fintdec$. 
The factor $\Fintdec$ takes values in $\Fintdec \in \{2,3,4,5,6,8,10,12,16,20,24,40\}$ for Gen~3 RFSoCs, and $\Fintdec \in \{2,4,8\}$ for Gen~1 and Gen~2 RFSoCs~\cite{amd_pg269}.
In most configurations, the 100\,GbE host-device link, instead of the data converters, limits the aggregate rate that can be transported between the devices and host.
For example, at $\Frfdc = 2.5\,\text{GS/s}$ and the minimum decimation factor $\Fintdec=2$, the ZCU216's 16 ADCs together produce 20\,{\gsps} of I/Q samples.
This corresponds to an aggregated I/Q sample payload rate of $\Rotw = 640\,\text{Gb/s}$ in \mycode{sc16} and is more than six times the capacity of the 100\,GbE link.
With one channel, the data converters set the limit: 1.25\,GS/s per channel on the ZCU216 and 2.5\,GS/s on the RFSoC~4x2. With two channels, the ZCU216 is still converter-limited at 1.25\,GS/s, while the RFSoC~4x2 is link-limited to 1.25\,GS/s in \mycode{sc16} and reaches 2.5\,GS/s only in \mycode{sc8}. With four or more channels, the 100\,GbE link becomes the limitation in every wire format.
As UHD does for a USRP, {\projectcode} coerces a requested I/Q sample rate to a supported one that fits the link in the selected wire format and reports the actual rate to the application.

\begin{figure}[!t]
    \centering
    \includegraphics[width=0.98\columnwidth]{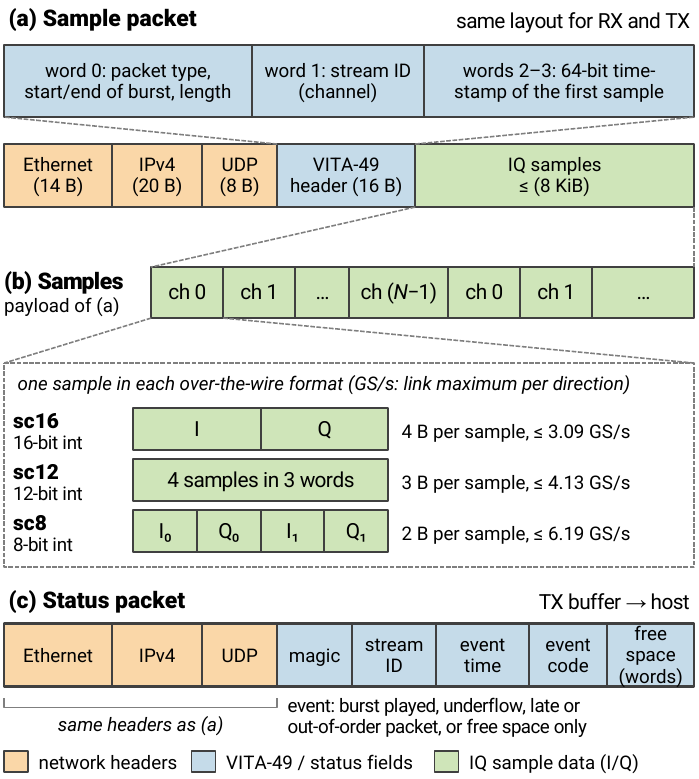}
    \vspace{-3.0mm}
    \caption{Packets over the 100\,GbE link.
    (a) A sample packet carries a VITA-49 header after Ethernet/IPv4/UDP.
    (b) Its payload interleaves the $N$ channels in one of three integer over-the-wire (OTW) formats: 16-, 12-, or 8-bit I and Q samples (sc16, sc12, sc8).
    (c) The buffer's status packet reports its free space and TX events.}
    \label{fig:packets}
    \vspace{-5.0mm}
\end{figure}

\myparatight{Wire format and resolution trade-off.}
The wire format is the representation of I/Q samples when transported over the host-device link. 
{\projectcode} supports three integer wire formats: 16-, 12-, or 8-bit I and Q, denoted by \mycode{sc16}, \mycode{sc12}, and \mycode{sc8}, respectively.
Each packet interleaves I/Q samples from the $N$ channels sample by sample, as illustrated in Fig.~\ref{fig:packets}. 
Two OTW converters, one in the FPGA (\S\ref{sec:design:fpga}) and one on the host (\S\ref{sec:design:host}), translate between the data converters' 16-bit samples and the selected wire format.
The application selects the wire format through UHD's standard stream argument \mycode{otw\_format}.
Fewer bytes per sample allow proportionally more samples to be carried over the same link: up to approximately 3.09\,GS/s in \mycode{sc16}, 4.13\,GS/s in \mycode{sc12}, and 6.19\,GS/s in \mycode{sc8} per direction, after accounting for the packet header overhead.
For the ZCU216, using \mycode{sc8} doubles the highest per-channel I/Q sample rate for 4, 8, and 16 channels, from 625, 312.5, and 156.25\,MS/s to 1250, 625, and 312.5\,MS/s, respectively, reaching an aggregated I/Q sample rate of 5.0\,GS/s.
However, retaining the eight most significant bits (MSBs) of I and Q trades sample resolution for bandwidth; ideal 8-bit quantization gives an SQNR of about 50\,dB for a full-scale sinusoid, while the application's signal levels and dynamic range determine whether this precision is sufficient.

\begin{figure}[!t]
    \centering
    \includegraphics[width=0.95\columnwidth]{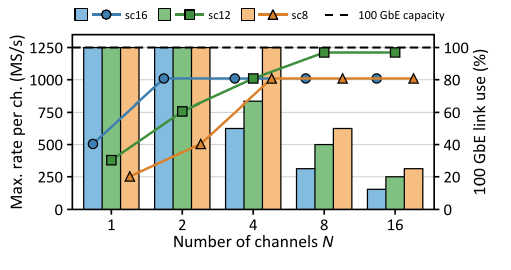}
    \vspace{-3.0mm}
    \caption{The highest supported per-channel I/Q sample rate for the ZCU216 board at $\Frfdc=2.5\,\text{GS/s}$ with \mycode{sc16}, \mycode{sc12}, and \mycode{sc8} wire formats (\emph{grouped bars, left axis}), and the corresponding 100\,{GbE} host-device link utilization (\emph{lines, right axis}).}
    \label{fig:link_usage}
    \vspace{-5.0mm}
\end{figure}

Fig.~\ref{fig:link_usage} summarizes this trade-off between per-channel bandwidth, channel count, and wire format on the ZCU216 at its maximum data converter sample rate, $\Frfdc = 2.5\,\text{GS/s}$.
For each channel count $N$, the bars show the highest per-channel I/Q sample rate in each wire format, and the lines show the usage of the 100\,GbE link.
The link usage is calculated from the payload rate $\Rotw$ defined in Table~\ref{tab:terms}.
For example, one \mycode{sc16} channel at 1.25\,{\gsps} ($\Fintdec=2$) per-channel needs $\Rotw = 1 \times 1.25\,\text{GS/s} \times 32\,\text{bit/sample} = 40\,\text{Gb/s}$.
Since $\Fintdec$ can only take the integer values listed above, $\Rotw$ changes in discrete steps without further resampling in the FPGA fabric.
Using the \mycode{sc16} wire format reaches its highest link usage with $N=2$ channels at 1.25\,{\gsps} per-channel ($\Fintdec=2$), where $\Rotw = 2 \times 1.25\,\text{GS/s} \times 32\,\text{bit/sample} = 80\,\text{Gb/s}$.
If the channel count is doubled to $N=4$, the next possible per-channel I/Q sample rate is 833.33\,{\msps} ($\Fintdec=3$), and it would exceed the link capacity, since $\Rotw = 4 \times 833.33\,\text{MS/s} \times 32\,\text{bit/sample} = 106.7\,\text{Gb/s}$.
Therefore, the highest rate that fits within the 100\,{GbE} link capacity is 625\,{\msps} ($\Fintdec=4$), which keeps $\Rotw = 80\,\text{Gb/s}$.
The same scenario also applies to the \mycode{sc8} wire format, so that the link utilization for \mycode{sc8} also plateaus as the channel count doubles in Fig.~\ref{fig:link_usage}.
Using \mycode{sc12} gets closer to the full link capacity at eight channels with 500\,{\msps} per-channel ($\Fintdec=5$) since $\Rotw = 8 \times 500\,\text{MS/s} \times 24\,\text{bit/sample} = 96\,\text{Gb/s}$.
Fig.~\ref{fig:link_rates} extends this calculation to every RFDC rate, wire format, channel count, and per-channel I/Q sample rate of the two boards, marking the rates that exceed the link.

\myparatight{Supporting 3GPP sample rates.}
I/Q sample rates such as 61.44\,{\msps} and 122.88\,{\msps}, used by LTE and 5G NR, cannot be obtained by integer decimation of RF data converter rates of 2.5\,{\gsps} or 5.0\,{\gsps}.
{\projectcode} also supports 3GPP-compliant RF data converter sample rate of $\Frfdc = 2.4576\,\text{GS/s}$.
With properly selected interpolation and decimation factor $\Fintdec$, per-channel I/Q sample rate from 1228.8\,MS/s down to 61.44\,MS/s can be effectively supported.
This includes 61.44\,MS/s, 122.88\,MS/s, 245.76\,MS/s, and 491.52\,MS/s, allowing standard 5G NR applications such as OCUDU~\cite{ocudu} (see \S\ref{sec:eval:ocudu}) to use these I/Q sample rates without resampling in the FPGA fabric or on the host.

\myparatight{Carrier frequency.}
The RF data converter's numerically controlled oscillator (NCO) sets the carrier frequency $\Fcarr$ within the supported first Nyquist zone: {0--1.25}\,GHz on the ZCU216 and {0--2.5}\,GHz on the RFSoC~4x2, or {0--1.2288}\,GHz and {0--2.4576}\,GHz with the respective 3GPP-compliant data converter sample rates.
This range can be extended using external RF front ends or operation in higher Nyquist zones, subject to the data converter's analog bandwidth.

\myparatight{Supported board configurations.}
Table~\ref{tab:rfsoc-board-config} lists the highest per-channel I/Q sample rate for each RFSoC board type, the data converter sample rate, channel count, and wire format.
Each board offers three data converter sample rates, and according to the $\Fsamp = \Frfdc/\Fintdec$ calculation, multiple sets of I/Q sample rates can be derived.
The highest, 2.5\,GS/s on the ZCU216 and 5.0\,GS/s on the RFSoC~4x2, is the maximum sample rate of the boards' ADCs and gives the largest per-channel bandwidth. 
The 2.4576\,GS/s and 4.9152\,GS/s rates are integer multiples of the 30.72\,MHz base rate of LTE and 5G NR. So their derived I/Q sample rates (e.g., 61.44, 122.88, 245.76, and 491.52\,MS/s) allow {\projectcode} to natively support standard LTE or 5G NR applications.
The 2.4\,GS/s and 4.8\,GS/s rates are integer multiples of the 20.0\,MHz Wi-Fi base rate, so their derived rates (e.g., 60, 120, and 240\,MS/s) allow native support for Wi-Fi applications in {\projectcode}.

\subsection{FPGA Data Plane}
\label{sec:design:fpga}

The FPGA data plane connects the RF data converters (DACs and ADCs) to the host over a single 100\,GbE link.
On the RX side, it merges the $N$ selected ADC channels into one interleaved stream, applies the phasers, and packetizes the I/Q samples.
On the TX side, it buffers the incoming stream, applies the phasers, and splits it to the $N$ DAC channels.
In both directions, an OTW converter module in the FPGA translates between the DACs/ADCs' 16-bit samples and the wire format selected by the application.

\myparatight{RX streamer (ADC$\rightarrow$host).}
The RX streamer passes captured I/Q samples through the OTW converter and executes UHD's timed stream commands.
It frames the samples as UDP packets shown in Fig.~\ref{fig:packets}(a) and stamps each packet with a 64-bit timestamp in a VITA-49-style header~\cite{vita49}.
The packets are spread over several UDP ports, so the host can receive them on parallel NIC queues.
In an overload scenario, the RX streamer drops whole packets, so any loss appears to the host as a timestamp gap and is reported rather than silently corrupting the stream.

\myparatight{TX streamer (host$\rightarrow$DAC).}
The TX streamer reverses the RX data path described above: it validates incoming packets from the host, holds the I/Q samples inside the 4\,MiB of on-chip memory, and starts each burst transmission at its scheduled time.
A second OTW converter restores 16-bit samples for the DACs, and the status packets shown in Fig.~\ref{fig:packets}(c) report burst completions, underflows, late starts, and sequence errors back to the host.
Because the DACs consume samples at a fixed rate, the host must neither overrun the buffer nor let it drain.
The streamer therefore reports periodically how many samples it has played, from which the host knows the buffer's free space, and the host sends only as many samples as fit.
Only half of the buffer is counted as free space, so the other half absorbs short host stalls.

\myparatight{Per-channel phasers.}
With two or more channels, the FPGA multiplies each channel by a complex-valued weight $w_n \in \mathbb{C}$ that the application sets through UHD at runtime.
On TX, the FPGA multiplies the stream prepared for the $n$-th DAC channel by $w_n$, a complex-valued weight stored as a pair of 16-bit fixed-point numbers; on RX, it multiplies the stream received from the $n$-th ADC channel by $w_n$. 
The weights serve two purposes: \emph{beamforming}, which steers the beam to a requested angle, and \emph{front-end calibration}, since external cables, amplifiers, and antennas add a fixed phase offset per chain that can be measured and calibrated out. 

\definecolor{lrU0}{HTML}{EEF5FB}
\definecolor{lrU1}{HTML}{DCE9F4}
\definecolor{lrU2}{HTML}{CBDCED}
\definecolor{lrU3}{HTML}{B9D0E6}
\definecolor{lrU4}{HTML}{A7C4DF}
\definecolor{lrU5}{HTML}{96B7D9}
\definecolor{lrU6}{HTML}{84ABD2}
\definecolor{lrU7}{HTML}{729FCB}
\definecolor{lrU8}{HTML}{6192C4}
\definecolor{lrU9}{HTML}{4F86BD}
\colorlet{lrOver}{fsLightRed}
\begin{figure*}[!t]
  \centering
  \fontsize{5.5}{6.6}\selectfont      %
  \setlength{\tabcolsep}{0.6pt}
  \vspace{-1.0mm}
  \begin{subfigure}{0pt}\phantomsubcaption\label{fig:link_rates:zcu216}\end{subfigure}%
  {\small\textbf{(\thesubfigure)~ZCU216}}\hfill\begin{tabular}{@{}l*{10}{>{\centering\arraybackslash}p{3mm}}c@{}}Link usage (\%) up to: & \cellcolor{lrU0}10 & \cellcolor{lrU1}20 & \cellcolor{lrU2}30 & \cellcolor{lrU3}40 & \cellcolor{lrU4}50 & \cellcolor{lrU5}60 & \cellcolor{lrU6}70 & \cellcolor{lrU7}80 & \cellcolor{lrU8}90 & \cellcolor{lrU9}100 & \cellcolor{lrOver}$>$100\end{tabular}\\[0.3mm]
  \begin{subfigure}[t]{0.325\textwidth}
    \centering
    \begin{tabularx}{\linewidth}{@{}r@{\hspace{1.5pt}}|*{12}{>{\raggedleft\arraybackslash}X}@{}}
        $N$ & \rotatebox{90}{62.5} & \rotatebox{90}{104.17} & \rotatebox{90}{125} & \rotatebox{90}{156.25} & \rotatebox{90}{208.33} & \rotatebox{90}{250} & \rotatebox{90}{312.5} & \rotatebox{90}{416.67} & \rotatebox{90}{500} & \rotatebox{90}{625} & \rotatebox{90}{833.33} & \rotatebox{90}{1250} \\
        \hline
        1 & \cellcolor{lrU0}2 & \cellcolor{lrU0}3.3 & \cellcolor{lrU0}4 & \cellcolor{lrU0}5 & \cellcolor{lrU0}6.7 & \cellcolor{lrU0}8 & \cellcolor{lrU1}10 & \cellcolor{lrU1}13.3 & \cellcolor{lrU1}16 & \cellcolor{lrU2}20 & \cellcolor{lrU2}26.7 & \cellcolor{lrU4}\textbf{40} \\
        2 & \cellcolor{lrU0}4 & \cellcolor{lrU0}6.7 & \cellcolor{lrU0}8 & \cellcolor{lrU1}10 & \cellcolor{lrU1}13.3 & \cellcolor{lrU1}16 & \cellcolor{lrU2}20 & \cellcolor{lrU2}26.7 & \cellcolor{lrU3}32 & \cellcolor{lrU4}40 & \cellcolor{lrU5}53.3 & \cellcolor{lrU8}\textbf{80} \\
        4 & \cellcolor{lrU0}8 & \cellcolor{lrU1}13.3 & \cellcolor{lrU1}16 & \cellcolor{lrU2}20 & \cellcolor{lrU2}26.7 & \cellcolor{lrU3}32 & \cellcolor{lrU4}40 & \cellcolor{lrU5}53.3 & \cellcolor{lrU6}64 & \cellcolor{lrU8}\textbf{80} & \cellcolor{lrOver}106.7 & \cellcolor{lrOver}160 \\
        8 & \cellcolor{lrU1}16 & \cellcolor{lrU2}26.7 & \cellcolor{lrU3}32 & \cellcolor{lrU4}40 & \cellcolor{lrU5}53.3 & \cellcolor{lrU6}64 & \cellcolor{lrU8}\textbf{80} & \cellcolor{lrOver}106.7 & \cellcolor{lrOver}128 & \cellcolor{lrOver}160 & \cellcolor{lrOver}213.3 & \cellcolor{lrOver}320 \\
        16 & \cellcolor{lrU3}32 & \cellcolor{lrU5}53.3 & \cellcolor{lrU6}64 & \cellcolor{lrU8}\textbf{80} & \cellcolor{lrOver}106.7 & \cellcolor{lrOver}128 & \cellcolor{lrOver}160 & \cellcolor{lrOver}213.3 & \cellcolor{lrOver}256 & \cellcolor{lrOver}320 & \cellcolor{lrOver}426.7 & \cellcolor{lrOver}640 \\
    \end{tabularx}
    \caption*{$\Frfdc = 2.5$\,GS/s, \mycode{sc16} (32\,bit)}
  \end{subfigure}\hfill
  \begin{subfigure}[t]{0.325\textwidth}
    \centering
    \begin{tabularx}{\linewidth}{@{}r@{\hspace{1.5pt}}|*{12}{>{\raggedleft\arraybackslash}X}@{}}
        $N$ & \rotatebox{90}{62.5} & \rotatebox{90}{104.17} & \rotatebox{90}{125} & \rotatebox{90}{156.25} & \rotatebox{90}{208.33} & \rotatebox{90}{250} & \rotatebox{90}{312.5} & \rotatebox{90}{416.67} & \rotatebox{90}{500} & \rotatebox{90}{625} & \rotatebox{90}{833.33} & \rotatebox{90}{1250} \\
        \hline
        1 & \cellcolor{lrU0}1.5 & \cellcolor{lrU0}2.5 & \cellcolor{lrU0}3 & \cellcolor{lrU0}3.8 & \cellcolor{lrU0}5 & \cellcolor{lrU0}6 & \cellcolor{lrU0}7.5 & \cellcolor{lrU1}10 & \cellcolor{lrU1}12 & \cellcolor{lrU1}15 & \cellcolor{lrU2}20 & \cellcolor{lrU3}\textbf{30} \\
        2 & \cellcolor{lrU0}3 & \cellcolor{lrU0}5 & \cellcolor{lrU0}6 & \cellcolor{lrU0}7.5 & \cellcolor{lrU1}10 & \cellcolor{lrU1}12 & \cellcolor{lrU1}15 & \cellcolor{lrU2}20 & \cellcolor{lrU2}24 & \cellcolor{lrU3}30 & \cellcolor{lrU4}40 & \cellcolor{lrU6}\textbf{60} \\
        4 & \cellcolor{lrU0}6 & \cellcolor{lrU1}10 & \cellcolor{lrU1}12 & \cellcolor{lrU1}15 & \cellcolor{lrU2}20 & \cellcolor{lrU2}24 & \cellcolor{lrU3}30 & \cellcolor{lrU4}40 & \cellcolor{lrU4}48 & \cellcolor{lrU6}60 & \cellcolor{lrU8}\textbf{80} & \cellcolor{lrOver}120 \\
        8 & \cellcolor{lrU1}12 & \cellcolor{lrU2}20 & \cellcolor{lrU2}24 & \cellcolor{lrU3}30 & \cellcolor{lrU4}40 & \cellcolor{lrU4}48 & \cellcolor{lrU6}60 & \cellcolor{lrU8}80 & \cellcolor{lrU9}\textbf{96} & \cellcolor{lrOver}120 & \cellcolor{lrOver}160 & \cellcolor{lrOver}240 \\
        16 & \cellcolor{lrU2}24 & \cellcolor{lrU4}40 & \cellcolor{lrU4}48 & \cellcolor{lrU6}60 & \cellcolor{lrU8}80 & \cellcolor{lrU9}\textbf{96} & \cellcolor{lrOver}120 & \cellcolor{lrOver}160 & \cellcolor{lrOver}192 & \cellcolor{lrOver}240 & \cellcolor{lrOver}320 & \cellcolor{lrOver}480 \\
    \end{tabularx}
    \caption*{$\Frfdc = 2.5$\,GS/s, \mycode{sc12} (24\,bit)}
  \end{subfigure}\hfill
  \begin{subfigure}[t]{0.325\textwidth}
    \centering
    \begin{tabularx}{\linewidth}{@{}r@{\hspace{1.5pt}}|*{12}{>{\raggedleft\arraybackslash}X}@{}}
        $N$ & \rotatebox{90}{62.5} & \rotatebox{90}{104.17} & \rotatebox{90}{125} & \rotatebox{90}{156.25} & \rotatebox{90}{208.33} & \rotatebox{90}{250} & \rotatebox{90}{312.5} & \rotatebox{90}{416.67} & \rotatebox{90}{500} & \rotatebox{90}{625} & \rotatebox{90}{833.33} & \rotatebox{90}{1250} \\
        \hline
        1 & \cellcolor{lrU0}1 & \cellcolor{lrU0}1.7 & \cellcolor{lrU0}2 & \cellcolor{lrU0}2.5 & \cellcolor{lrU0}3.3 & \cellcolor{lrU0}4 & \cellcolor{lrU0}5 & \cellcolor{lrU0}6.7 & \cellcolor{lrU0}8 & \cellcolor{lrU1}10 & \cellcolor{lrU1}13.3 & \cellcolor{lrU2}\textbf{20} \\
        2 & \cellcolor{lrU0}2 & \cellcolor{lrU0}3.3 & \cellcolor{lrU0}4 & \cellcolor{lrU0}5 & \cellcolor{lrU0}6.7 & \cellcolor{lrU0}8 & \cellcolor{lrU1}10 & \cellcolor{lrU1}13.3 & \cellcolor{lrU1}16 & \cellcolor{lrU2}20 & \cellcolor{lrU2}26.7 & \cellcolor{lrU4}\textbf{40} \\
        4 & \cellcolor{lrU0}4 & \cellcolor{lrU0}6.7 & \cellcolor{lrU0}8 & \cellcolor{lrU1}10 & \cellcolor{lrU1}13.3 & \cellcolor{lrU1}16 & \cellcolor{lrU2}20 & \cellcolor{lrU2}26.7 & \cellcolor{lrU3}32 & \cellcolor{lrU4}40 & \cellcolor{lrU5}53.3 & \cellcolor{lrU8}\textbf{80} \\
        8 & \cellcolor{lrU0}8 & \cellcolor{lrU1}13.3 & \cellcolor{lrU1}16 & \cellcolor{lrU2}20 & \cellcolor{lrU2}26.7 & \cellcolor{lrU3}32 & \cellcolor{lrU4}40 & \cellcolor{lrU5}53.3 & \cellcolor{lrU6}64 & \cellcolor{lrU8}\textbf{80} & \cellcolor{lrOver}106.7 & \cellcolor{lrOver}160 \\
        16 & \cellcolor{lrU1}16 & \cellcolor{lrU2}26.7 & \cellcolor{lrU3}32 & \cellcolor{lrU4}40 & \cellcolor{lrU5}53.3 & \cellcolor{lrU6}64 & \cellcolor{lrU8}\textbf{80} & \cellcolor{lrOver}106.7 & \cellcolor{lrOver}128 & \cellcolor{lrOver}160 & \cellcolor{lrOver}213.3 & \cellcolor{lrOver}320 \\
    \end{tabularx}
    \caption*{$\Frfdc = 2.5$\,GS/s, \mycode{sc8} (16\,bit)}
  \end{subfigure}\\[0.6mm]
  \begin{subfigure}[t]{0.325\textwidth}
    \centering
    \begin{tabularx}{\linewidth}{@{}r@{\hspace{1.5pt}}|*{12}{>{\raggedleft\arraybackslash}X}@{}}
        $N$ & \rotatebox{90}{61.44} & \rotatebox{90}{102.4} & \rotatebox{90}{122.88} & \rotatebox{90}{153.6} & \rotatebox{90}{204.8} & \rotatebox{90}{245.76} & \rotatebox{90}{307.2} & \rotatebox{90}{409.6} & \rotatebox{90}{491.52} & \rotatebox{90}{614.4} & \rotatebox{90}{819.2} & \rotatebox{90}{1228.8} \\
        \hline
        1 & \cellcolor{lrU0}2 & \cellcolor{lrU0}3.3 & \cellcolor{lrU0}3.9 & \cellcolor{lrU0}4.9 & \cellcolor{lrU0}6.6 & \cellcolor{lrU0}7.9 & \cellcolor{lrU0}9.8 & \cellcolor{lrU1}13.1 & \cellcolor{lrU1}15.7 & \cellcolor{lrU1}19.7 & \cellcolor{lrU2}26.2 & \cellcolor{lrU3}\textbf{39.3} \\
        2 & \cellcolor{lrU0}3.9 & \cellcolor{lrU0}6.6 & \cellcolor{lrU0}7.9 & \cellcolor{lrU0}9.8 & \cellcolor{lrU1}13.1 & \cellcolor{lrU1}15.7 & \cellcolor{lrU1}19.7 & \cellcolor{lrU2}26.2 & \cellcolor{lrU3}31.5 & \cellcolor{lrU3}39.3 & \cellcolor{lrU5}52.4 & \cellcolor{lrU7}\textbf{78.6} \\
        4 & \cellcolor{lrU0}7.9 & \cellcolor{lrU1}13.1 & \cellcolor{lrU1}15.7 & \cellcolor{lrU1}19.7 & \cellcolor{lrU2}26.2 & \cellcolor{lrU3}31.5 & \cellcolor{lrU3}39.3 & \cellcolor{lrU5}52.4 & \cellcolor{lrU6}62.9 & \cellcolor{lrU7}\textbf{78.6} & \cellcolor{lrOver}104.9 & \cellcolor{lrOver}157.3 \\
        8 & \cellcolor{lrU1}15.7 & \cellcolor{lrU2}26.2 & \cellcolor{lrU3}31.5 & \cellcolor{lrU3}39.3 & \cellcolor{lrU5}52.4 & \cellcolor{lrU6}62.9 & \cellcolor{lrU7}\textbf{78.6} & \cellcolor{lrOver}104.9 & \cellcolor{lrOver}125.8 & \cellcolor{lrOver}157.3 & \cellcolor{lrOver}209.7 & \cellcolor{lrOver}314.6 \\
        16 & \cellcolor{lrU3}31.5 & \cellcolor{lrU5}52.4 & \cellcolor{lrU6}62.9 & \cellcolor{lrU7}\textbf{78.6} & \cellcolor{lrOver}104.9 & \cellcolor{lrOver}125.8 & \cellcolor{lrOver}157.3 & \cellcolor{lrOver}209.7 & \cellcolor{lrOver}251.7 & \cellcolor{lrOver}314.6 & \cellcolor{lrOver}419.4 & \cellcolor{lrOver}629.1 \\
    \end{tabularx}
    \caption*{$\Frfdc = 2.4576$\,GS/s, \mycode{sc16} (32\,bit)}
  \end{subfigure}\hfill
  \begin{subfigure}[t]{0.325\textwidth}
    \centering
    \begin{tabularx}{\linewidth}{@{}r@{\hspace{1.5pt}}|*{12}{>{\raggedleft\arraybackslash}X}@{}}
        $N$ & \rotatebox{90}{61.44} & \rotatebox{90}{102.4} & \rotatebox{90}{122.88} & \rotatebox{90}{153.6} & \rotatebox{90}{204.8} & \rotatebox{90}{245.76} & \rotatebox{90}{307.2} & \rotatebox{90}{409.6} & \rotatebox{90}{491.52} & \rotatebox{90}{614.4} & \rotatebox{90}{819.2} & \rotatebox{90}{1228.8} \\
        \hline
        1 & \cellcolor{lrU0}1.5 & \cellcolor{lrU0}2.5 & \cellcolor{lrU0}2.9 & \cellcolor{lrU0}3.7 & \cellcolor{lrU0}4.9 & \cellcolor{lrU0}5.9 & \cellcolor{lrU0}7.4 & \cellcolor{lrU0}9.8 & \cellcolor{lrU1}11.8 & \cellcolor{lrU1}14.7 & \cellcolor{lrU1}19.7 & \cellcolor{lrU2}\textbf{29.5} \\
        2 & \cellcolor{lrU0}2.9 & \cellcolor{lrU0}4.9 & \cellcolor{lrU0}5.9 & \cellcolor{lrU0}7.4 & \cellcolor{lrU0}9.8 & \cellcolor{lrU1}11.8 & \cellcolor{lrU1}14.7 & \cellcolor{lrU1}19.7 & \cellcolor{lrU2}23.6 & \cellcolor{lrU2}29.5 & \cellcolor{lrU3}39.3 & \cellcolor{lrU5}\textbf{59} \\
        4 & \cellcolor{lrU0}5.9 & \cellcolor{lrU0}9.8 & \cellcolor{lrU1}11.8 & \cellcolor{lrU1}14.7 & \cellcolor{lrU1}19.7 & \cellcolor{lrU2}23.6 & \cellcolor{lrU2}29.5 & \cellcolor{lrU3}39.3 & \cellcolor{lrU4}47.2 & \cellcolor{lrU5}59 & \cellcolor{lrU7}\textbf{78.6} & \cellcolor{lrOver}118 \\
        8 & \cellcolor{lrU1}11.8 & \cellcolor{lrU1}19.7 & \cellcolor{lrU2}23.6 & \cellcolor{lrU2}29.5 & \cellcolor{lrU3}39.3 & \cellcolor{lrU4}47.2 & \cellcolor{lrU5}59 & \cellcolor{lrU7}78.6 & \cellcolor{lrU9}\textbf{94.4} & \cellcolor{lrOver}118 & \cellcolor{lrOver}157.3 & \cellcolor{lrOver}235.9 \\
        16 & \cellcolor{lrU2}23.6 & \cellcolor{lrU3}39.3 & \cellcolor{lrU4}47.2 & \cellcolor{lrU5}59 & \cellcolor{lrU7}78.6 & \cellcolor{lrU9}\textbf{94.4} & \cellcolor{lrOver}118 & \cellcolor{lrOver}157.3 & \cellcolor{lrOver}188.7 & \cellcolor{lrOver}235.9 & \cellcolor{lrOver}314.6 & \cellcolor{lrOver}471.9 \\
    \end{tabularx}
    \caption*{$\Frfdc = 2.4576$\,GS/s, \mycode{sc12} (24\,bit)}
  \end{subfigure}\hfill
  \begin{subfigure}[t]{0.325\textwidth}
    \centering
    \begin{tabularx}{\linewidth}{@{}r@{\hspace{1.5pt}}|*{12}{>{\raggedleft\arraybackslash}X}@{}}
        $N$ & \rotatebox{90}{61.44} & \rotatebox{90}{102.4} & \rotatebox{90}{122.88} & \rotatebox{90}{153.6} & \rotatebox{90}{204.8} & \rotatebox{90}{245.76} & \rotatebox{90}{307.2} & \rotatebox{90}{409.6} & \rotatebox{90}{491.52} & \rotatebox{90}{614.4} & \rotatebox{90}{819.2} & \rotatebox{90}{1228.8} \\
        \hline
        1 & \cellcolor{lrU0}1 & \cellcolor{lrU0}1.6 & \cellcolor{lrU0}2 & \cellcolor{lrU0}2.5 & \cellcolor{lrU0}3.3 & \cellcolor{lrU0}3.9 & \cellcolor{lrU0}4.9 & \cellcolor{lrU0}6.6 & \cellcolor{lrU0}7.9 & \cellcolor{lrU0}9.8 & \cellcolor{lrU1}13.1 & \cellcolor{lrU1}\textbf{19.7} \\
        2 & \cellcolor{lrU0}2 & \cellcolor{lrU0}3.3 & \cellcolor{lrU0}3.9 & \cellcolor{lrU0}4.9 & \cellcolor{lrU0}6.6 & \cellcolor{lrU0}7.9 & \cellcolor{lrU0}9.8 & \cellcolor{lrU1}13.1 & \cellcolor{lrU1}15.7 & \cellcolor{lrU1}19.7 & \cellcolor{lrU2}26.2 & \cellcolor{lrU3}\textbf{39.3} \\
        4 & \cellcolor{lrU0}3.9 & \cellcolor{lrU0}6.6 & \cellcolor{lrU0}7.9 & \cellcolor{lrU0}9.8 & \cellcolor{lrU1}13.1 & \cellcolor{lrU1}15.7 & \cellcolor{lrU1}19.7 & \cellcolor{lrU2}26.2 & \cellcolor{lrU3}31.5 & \cellcolor{lrU3}39.3 & \cellcolor{lrU5}52.4 & \cellcolor{lrU7}\textbf{78.6} \\
        8 & \cellcolor{lrU0}7.9 & \cellcolor{lrU1}13.1 & \cellcolor{lrU1}15.7 & \cellcolor{lrU1}19.7 & \cellcolor{lrU2}26.2 & \cellcolor{lrU3}31.5 & \cellcolor{lrU3}39.3 & \cellcolor{lrU5}52.4 & \cellcolor{lrU6}62.9 & \cellcolor{lrU7}\textbf{78.6} & \cellcolor{lrOver}104.9 & \cellcolor{lrOver}157.3 \\
        16 & \cellcolor{lrU1}15.7 & \cellcolor{lrU2}26.2 & \cellcolor{lrU3}31.5 & \cellcolor{lrU3}39.3 & \cellcolor{lrU5}52.4 & \cellcolor{lrU6}62.9 & \cellcolor{lrU7}\textbf{78.6} & \cellcolor{lrOver}104.9 & \cellcolor{lrOver}125.8 & \cellcolor{lrOver}157.3 & \cellcolor{lrOver}209.7 & \cellcolor{lrOver}314.6 \\
    \end{tabularx}
    \caption*{$\Frfdc = 2.4576$\,GS/s, \mycode{sc8} (16\,bit)}
  \end{subfigure}\\[0.6mm]
  \begin{subfigure}[t]{0.325\textwidth}
    \centering
    \begin{tabularx}{\linewidth}{@{}r@{\hspace{1.5pt}}|*{12}{>{\raggedleft\arraybackslash}X}@{}}
        $N$ & \rotatebox{90}{60} & \rotatebox{90}{100} & \rotatebox{90}{120} & \rotatebox{90}{150} & \rotatebox{90}{200} & \rotatebox{90}{240} & \rotatebox{90}{300} & \rotatebox{90}{400} & \rotatebox{90}{480} & \rotatebox{90}{600} & \rotatebox{90}{800} & \rotatebox{90}{1200} \\
        \hline
        1 & \cellcolor{lrU0}1.9 & \cellcolor{lrU0}3.2 & \cellcolor{lrU0}3.8 & \cellcolor{lrU0}4.8 & \cellcolor{lrU0}6.4 & \cellcolor{lrU0}7.7 & \cellcolor{lrU0}9.6 & \cellcolor{lrU1}12.8 & \cellcolor{lrU1}15.4 & \cellcolor{lrU1}19.2 & \cellcolor{lrU2}25.6 & \cellcolor{lrU3}\textbf{38.4} \\
        2 & \cellcolor{lrU0}3.8 & \cellcolor{lrU0}6.4 & \cellcolor{lrU0}7.7 & \cellcolor{lrU0}9.6 & \cellcolor{lrU1}12.8 & \cellcolor{lrU1}15.4 & \cellcolor{lrU1}19.2 & \cellcolor{lrU2}25.6 & \cellcolor{lrU3}30.7 & \cellcolor{lrU3}38.4 & \cellcolor{lrU5}51.2 & \cellcolor{lrU7}\textbf{76.8} \\
        4 & \cellcolor{lrU0}7.7 & \cellcolor{lrU1}12.8 & \cellcolor{lrU1}15.4 & \cellcolor{lrU1}19.2 & \cellcolor{lrU2}25.6 & \cellcolor{lrU3}30.7 & \cellcolor{lrU3}38.4 & \cellcolor{lrU5}51.2 & \cellcolor{lrU6}61.4 & \cellcolor{lrU7}\textbf{76.8} & \cellcolor{lrOver}102.4 & \cellcolor{lrOver}153.6 \\
        8 & \cellcolor{lrU1}15.4 & \cellcolor{lrU2}25.6 & \cellcolor{lrU3}30.7 & \cellcolor{lrU3}38.4 & \cellcolor{lrU5}51.2 & \cellcolor{lrU6}61.4 & \cellcolor{lrU7}\textbf{76.8} & \cellcolor{lrOver}102.4 & \cellcolor{lrOver}122.9 & \cellcolor{lrOver}153.6 & \cellcolor{lrOver}204.8 & \cellcolor{lrOver}307.2 \\
        16 & \cellcolor{lrU3}30.7 & \cellcolor{lrU5}51.2 & \cellcolor{lrU6}61.4 & \cellcolor{lrU7}\textbf{76.8} & \cellcolor{lrOver}102.4 & \cellcolor{lrOver}122.9 & \cellcolor{lrOver}153.6 & \cellcolor{lrOver}204.8 & \cellcolor{lrOver}245.8 & \cellcolor{lrOver}307.2 & \cellcolor{lrOver}409.6 & \cellcolor{lrOver}614.4 \\
    \end{tabularx}
    \caption*{$\Frfdc = 2.4$\,GS/s, \mycode{sc16} (32\,bit)}
  \end{subfigure}\hfill
  \begin{subfigure}[t]{0.325\textwidth}
    \centering
    \begin{tabularx}{\linewidth}{@{}r@{\hspace{1.5pt}}|*{12}{>{\raggedleft\arraybackslash}X}@{}}
        $N$ & \rotatebox{90}{60} & \rotatebox{90}{100} & \rotatebox{90}{120} & \rotatebox{90}{150} & \rotatebox{90}{200} & \rotatebox{90}{240} & \rotatebox{90}{300} & \rotatebox{90}{400} & \rotatebox{90}{480} & \rotatebox{90}{600} & \rotatebox{90}{800} & \rotatebox{90}{1200} \\
        \hline
        1 & \cellcolor{lrU0}1.4 & \cellcolor{lrU0}2.4 & \cellcolor{lrU0}2.9 & \cellcolor{lrU0}3.6 & \cellcolor{lrU0}4.8 & \cellcolor{lrU0}5.8 & \cellcolor{lrU0}7.2 & \cellcolor{lrU0}9.6 & \cellcolor{lrU1}11.5 & \cellcolor{lrU1}14.4 & \cellcolor{lrU1}19.2 & \cellcolor{lrU2}\textbf{28.8} \\
        2 & \cellcolor{lrU0}2.9 & \cellcolor{lrU0}4.8 & \cellcolor{lrU0}5.8 & \cellcolor{lrU0}7.2 & \cellcolor{lrU0}9.6 & \cellcolor{lrU1}11.5 & \cellcolor{lrU1}14.4 & \cellcolor{lrU1}19.2 & \cellcolor{lrU2}23 & \cellcolor{lrU2}28.8 & \cellcolor{lrU3}38.4 & \cellcolor{lrU5}\textbf{57.6} \\
        4 & \cellcolor{lrU0}5.8 & \cellcolor{lrU0}9.6 & \cellcolor{lrU1}11.5 & \cellcolor{lrU1}14.4 & \cellcolor{lrU1}19.2 & \cellcolor{lrU2}23 & \cellcolor{lrU2}28.8 & \cellcolor{lrU3}38.4 & \cellcolor{lrU4}46.1 & \cellcolor{lrU5}57.6 & \cellcolor{lrU7}\textbf{76.8} & \cellcolor{lrOver}115.2 \\
        8 & \cellcolor{lrU1}11.5 & \cellcolor{lrU1}19.2 & \cellcolor{lrU2}23 & \cellcolor{lrU2}28.8 & \cellcolor{lrU3}38.4 & \cellcolor{lrU4}46.1 & \cellcolor{lrU5}57.6 & \cellcolor{lrU7}76.8 & \cellcolor{lrU9}\textbf{92.2} & \cellcolor{lrOver}115.2 & \cellcolor{lrOver}153.6 & \cellcolor{lrOver}230.4 \\
        16 & \cellcolor{lrU2}23 & \cellcolor{lrU3}38.4 & \cellcolor{lrU4}46.1 & \cellcolor{lrU5}57.6 & \cellcolor{lrU7}76.8 & \cellcolor{lrU9}\textbf{92.2} & \cellcolor{lrOver}115.2 & \cellcolor{lrOver}153.6 & \cellcolor{lrOver}184.3 & \cellcolor{lrOver}230.4 & \cellcolor{lrOver}307.2 & \cellcolor{lrOver}460.8 \\
    \end{tabularx}
    \caption*{$\Frfdc = 2.4$\,GS/s, \mycode{sc12} (24\,bit)}
  \end{subfigure}\hfill
  \begin{subfigure}[t]{0.325\textwidth}
    \centering
    \begin{tabularx}{\linewidth}{@{}r@{\hspace{1.5pt}}|*{12}{>{\raggedleft\arraybackslash}X}@{}}
        $N$ & \rotatebox{90}{60} & \rotatebox{90}{100} & \rotatebox{90}{120} & \rotatebox{90}{150} & \rotatebox{90}{200} & \rotatebox{90}{240} & \rotatebox{90}{300} & \rotatebox{90}{400} & \rotatebox{90}{480} & \rotatebox{90}{600} & \rotatebox{90}{800} & \rotatebox{90}{1200} \\
        \hline
        1 & \cellcolor{lrU0}1 & \cellcolor{lrU0}1.6 & \cellcolor{lrU0}1.9 & \cellcolor{lrU0}2.4 & \cellcolor{lrU0}3.2 & \cellcolor{lrU0}3.8 & \cellcolor{lrU0}4.8 & \cellcolor{lrU0}6.4 & \cellcolor{lrU0}7.7 & \cellcolor{lrU0}9.6 & \cellcolor{lrU1}12.8 & \cellcolor{lrU1}\textbf{19.2} \\
        2 & \cellcolor{lrU0}1.9 & \cellcolor{lrU0}3.2 & \cellcolor{lrU0}3.8 & \cellcolor{lrU0}4.8 & \cellcolor{lrU0}6.4 & \cellcolor{lrU0}7.7 & \cellcolor{lrU0}9.6 & \cellcolor{lrU1}12.8 & \cellcolor{lrU1}15.4 & \cellcolor{lrU1}19.2 & \cellcolor{lrU2}25.6 & \cellcolor{lrU3}\textbf{38.4} \\
        4 & \cellcolor{lrU0}3.8 & \cellcolor{lrU0}6.4 & \cellcolor{lrU0}7.7 & \cellcolor{lrU0}9.6 & \cellcolor{lrU1}12.8 & \cellcolor{lrU1}15.4 & \cellcolor{lrU1}19.2 & \cellcolor{lrU2}25.6 & \cellcolor{lrU3}30.7 & \cellcolor{lrU3}38.4 & \cellcolor{lrU5}51.2 & \cellcolor{lrU7}\textbf{76.8} \\
        8 & \cellcolor{lrU0}7.7 & \cellcolor{lrU1}12.8 & \cellcolor{lrU1}15.4 & \cellcolor{lrU1}19.2 & \cellcolor{lrU2}25.6 & \cellcolor{lrU3}30.7 & \cellcolor{lrU3}38.4 & \cellcolor{lrU5}51.2 & \cellcolor{lrU6}61.4 & \cellcolor{lrU7}\textbf{76.8} & \cellcolor{lrOver}102.4 & \cellcolor{lrOver}153.6 \\
        16 & \cellcolor{lrU1}15.4 & \cellcolor{lrU2}25.6 & \cellcolor{lrU3}30.7 & \cellcolor{lrU3}38.4 & \cellcolor{lrU5}51.2 & \cellcolor{lrU6}61.4 & \cellcolor{lrU7}\textbf{76.8} & \cellcolor{lrOver}102.4 & \cellcolor{lrOver}122.9 & \cellcolor{lrOver}153.6 & \cellcolor{lrOver}204.8 & \cellcolor{lrOver}307.2 \\
    \end{tabularx}
    \caption*{$\Frfdc = 2.4$\,GS/s, \mycode{sc8} (16\,bit)}
  \end{subfigure}\\[0.6mm]
  \begin{subfigure}{0pt}\phantomsubcaption\label{fig:link_rates:rfsoc4x2}\end{subfigure}%
  {\small\textbf{(\thesubfigure)~RFSoC~4x2}}\hfill\mbox{}\\[0.3mm]
  \begin{subfigure}[t]{0.325\textwidth}
    \centering
    \begin{tabularx}{\linewidth}{@{}r@{\hspace{1.5pt}}|*{12}{>{\raggedleft\arraybackslash}X}@{}}
        $N$ & \rotatebox{90}{125} & \rotatebox{90}{208.33} & \rotatebox{90}{250} & \rotatebox{90}{312.5} & \rotatebox{90}{416.67} & \rotatebox{90}{500} & \rotatebox{90}{625} & \rotatebox{90}{833.33} & \rotatebox{90}{1000} & \rotatebox{90}{1250} & \rotatebox{90}{1666.67} & \rotatebox{90}{2500} \\
        \hline
        1 & \cellcolor{lrU0}4 & \cellcolor{lrU0}6.7 & \cellcolor{lrU0}8 & \cellcolor{lrU1}10 & \cellcolor{lrU1}13.3 & \cellcolor{lrU1}16 & \cellcolor{lrU2}20 & \cellcolor{lrU2}26.7 & \cellcolor{lrU3}32 & \cellcolor{lrU4}40 & \cellcolor{lrU5}53.3 & \cellcolor{lrU8}\textbf{80} \\
        2 & \cellcolor{lrU0}8 & \cellcolor{lrU1}13.3 & \cellcolor{lrU1}16 & \cellcolor{lrU2}20 & \cellcolor{lrU2}26.7 & \cellcolor{lrU3}32 & \cellcolor{lrU4}40 & \cellcolor{lrU5}53.3 & \cellcolor{lrU6}64 & \cellcolor{lrU8}\textbf{80} & \cellcolor{lrOver}106.7 & \cellcolor{lrOver}160 \\
    \end{tabularx}
    \caption*{$\Frfdc = 5.0$\,GS/s, \mycode{sc16} (32\,bit)}
  \end{subfigure}\hfill
  \begin{subfigure}[t]{0.325\textwidth}
    \centering
    \begin{tabularx}{\linewidth}{@{}r@{\hspace{1.5pt}}|*{12}{>{\raggedleft\arraybackslash}X}@{}}
        $N$ & \rotatebox{90}{125} & \rotatebox{90}{208.33} & \rotatebox{90}{250} & \rotatebox{90}{312.5} & \rotatebox{90}{416.67} & \rotatebox{90}{500} & \rotatebox{90}{625} & \rotatebox{90}{833.33} & \rotatebox{90}{1000} & \rotatebox{90}{1250} & \rotatebox{90}{1666.67} & \rotatebox{90}{2500} \\
        \hline
        1 & \cellcolor{lrU0}3 & \cellcolor{lrU0}5 & \cellcolor{lrU0}6 & \cellcolor{lrU0}7.5 & \cellcolor{lrU1}10 & \cellcolor{lrU1}12 & \cellcolor{lrU1}15 & \cellcolor{lrU2}20 & \cellcolor{lrU2}24 & \cellcolor{lrU3}30 & \cellcolor{lrU4}40 & \cellcolor{lrU6}\textbf{60} \\
        2 & \cellcolor{lrU0}6 & \cellcolor{lrU1}10 & \cellcolor{lrU1}12 & \cellcolor{lrU1}15 & \cellcolor{lrU2}20 & \cellcolor{lrU2}24 & \cellcolor{lrU3}30 & \cellcolor{lrU4}40 & \cellcolor{lrU4}48 & \cellcolor{lrU6}60 & \cellcolor{lrU8}\textbf{80} & \cellcolor{lrOver}120 \\
    \end{tabularx}
    \caption*{$\Frfdc = 5.0$\,GS/s, \mycode{sc12} (24\,bit)}
  \end{subfigure}\hfill
  \begin{subfigure}[t]{0.325\textwidth}
    \centering
    \begin{tabularx}{\linewidth}{@{}r@{\hspace{1.5pt}}|*{12}{>{\raggedleft\arraybackslash}X}@{}}
        $N$ & \rotatebox{90}{125} & \rotatebox{90}{208.33} & \rotatebox{90}{250} & \rotatebox{90}{312.5} & \rotatebox{90}{416.67} & \rotatebox{90}{500} & \rotatebox{90}{625} & \rotatebox{90}{833.33} & \rotatebox{90}{1000} & \rotatebox{90}{1250} & \rotatebox{90}{1666.67} & \rotatebox{90}{2500} \\
        \hline
        1 & \cellcolor{lrU0}2 & \cellcolor{lrU0}3.3 & \cellcolor{lrU0}4 & \cellcolor{lrU0}5 & \cellcolor{lrU0}6.7 & \cellcolor{lrU0}8 & \cellcolor{lrU1}10 & \cellcolor{lrU1}13.3 & \cellcolor{lrU1}16 & \cellcolor{lrU2}20 & \cellcolor{lrU2}26.7 & \cellcolor{lrU4}\textbf{40} \\
        2 & \cellcolor{lrU0}4 & \cellcolor{lrU0}6.7 & \cellcolor{lrU0}8 & \cellcolor{lrU1}10 & \cellcolor{lrU1}13.3 & \cellcolor{lrU1}16 & \cellcolor{lrU2}20 & \cellcolor{lrU2}26.7 & \cellcolor{lrU3}32 & \cellcolor{lrU4}40 & \cellcolor{lrU5}53.3 & \cellcolor{lrU8}\textbf{80} \\
    \end{tabularx}
    \caption*{$\Frfdc = 5.0$\,GS/s, \mycode{sc8} (16\,bit)}
  \end{subfigure}\\[0.6mm]
  \begin{subfigure}[t]{0.325\textwidth}
    \centering
    \begin{tabularx}{\linewidth}{@{}r@{\hspace{1.5pt}}|*{12}{>{\raggedleft\arraybackslash}X}@{}}
        $N$ & \rotatebox{90}{122.88} & \rotatebox{90}{204.8} & \rotatebox{90}{245.76} & \rotatebox{90}{307.2} & \rotatebox{90}{409.6} & \rotatebox{90}{491.52} & \rotatebox{90}{614.4} & \rotatebox{90}{819.2} & \rotatebox{90}{983.04} & \rotatebox{90}{1228.8} & \rotatebox{90}{1638.4} & \rotatebox{90}{2457.6} \\
        \hline
        1 & \cellcolor{lrU0}3.9 & \cellcolor{lrU0}6.6 & \cellcolor{lrU0}7.9 & \cellcolor{lrU0}9.8 & \cellcolor{lrU1}13.1 & \cellcolor{lrU1}15.7 & \cellcolor{lrU1}19.7 & \cellcolor{lrU2}26.2 & \cellcolor{lrU3}31.5 & \cellcolor{lrU3}39.3 & \cellcolor{lrU5}52.4 & \cellcolor{lrU7}\textbf{78.6} \\
        2 & \cellcolor{lrU0}7.9 & \cellcolor{lrU1}13.1 & \cellcolor{lrU1}15.7 & \cellcolor{lrU1}19.7 & \cellcolor{lrU2}26.2 & \cellcolor{lrU3}31.5 & \cellcolor{lrU3}39.3 & \cellcolor{lrU5}52.4 & \cellcolor{lrU6}62.9 & \cellcolor{lrU7}\textbf{78.6} & \cellcolor{lrOver}104.9 & \cellcolor{lrOver}157.3 \\
    \end{tabularx}
    \caption*{$\Frfdc = 4.9152$\,GS/s, \mycode{sc16} (32\,bit)}
  \end{subfigure}\hfill
  \begin{subfigure}[t]{0.325\textwidth}
    \centering
    \begin{tabularx}{\linewidth}{@{}r@{\hspace{1.5pt}}|*{12}{>{\raggedleft\arraybackslash}X}@{}}
        $N$ & \rotatebox{90}{122.88} & \rotatebox{90}{204.8} & \rotatebox{90}{245.76} & \rotatebox{90}{307.2} & \rotatebox{90}{409.6} & \rotatebox{90}{491.52} & \rotatebox{90}{614.4} & \rotatebox{90}{819.2} & \rotatebox{90}{983.04} & \rotatebox{90}{1228.8} & \rotatebox{90}{1638.4} & \rotatebox{90}{2457.6} \\
        \hline
        1 & \cellcolor{lrU0}2.9 & \cellcolor{lrU0}4.9 & \cellcolor{lrU0}5.9 & \cellcolor{lrU0}7.4 & \cellcolor{lrU0}9.8 & \cellcolor{lrU1}11.8 & \cellcolor{lrU1}14.7 & \cellcolor{lrU1}19.7 & \cellcolor{lrU2}23.6 & \cellcolor{lrU2}29.5 & \cellcolor{lrU3}39.3 & \cellcolor{lrU5}\textbf{59} \\
        2 & \cellcolor{lrU0}5.9 & \cellcolor{lrU0}9.8 & \cellcolor{lrU1}11.8 & \cellcolor{lrU1}14.7 & \cellcolor{lrU1}19.7 & \cellcolor{lrU2}23.6 & \cellcolor{lrU2}29.5 & \cellcolor{lrU3}39.3 & \cellcolor{lrU4}47.2 & \cellcolor{lrU5}59 & \cellcolor{lrU7}\textbf{78.6} & \cellcolor{lrOver}118 \\
    \end{tabularx}
    \caption*{$\Frfdc = 4.9152$\,GS/s, \mycode{sc12} (24\,bit)}
  \end{subfigure}\hfill
  \begin{subfigure}[t]{0.325\textwidth}
    \centering
    \begin{tabularx}{\linewidth}{@{}r@{\hspace{1.5pt}}|*{12}{>{\raggedleft\arraybackslash}X}@{}}
        $N$ & \rotatebox{90}{122.88} & \rotatebox{90}{204.8} & \rotatebox{90}{245.76} & \rotatebox{90}{307.2} & \rotatebox{90}{409.6} & \rotatebox{90}{491.52} & \rotatebox{90}{614.4} & \rotatebox{90}{819.2} & \rotatebox{90}{983.04} & \rotatebox{90}{1228.8} & \rotatebox{90}{1638.4} & \rotatebox{90}{2457.6} \\
        \hline
        1 & \cellcolor{lrU0}2 & \cellcolor{lrU0}3.3 & \cellcolor{lrU0}3.9 & \cellcolor{lrU0}4.9 & \cellcolor{lrU0}6.6 & \cellcolor{lrU0}7.9 & \cellcolor{lrU0}9.8 & \cellcolor{lrU1}13.1 & \cellcolor{lrU1}15.7 & \cellcolor{lrU1}19.7 & \cellcolor{lrU2}26.2 & \cellcolor{lrU3}\textbf{39.3} \\
        2 & \cellcolor{lrU0}3.9 & \cellcolor{lrU0}6.6 & \cellcolor{lrU0}7.9 & \cellcolor{lrU0}9.8 & \cellcolor{lrU1}13.1 & \cellcolor{lrU1}15.7 & \cellcolor{lrU1}19.7 & \cellcolor{lrU2}26.2 & \cellcolor{lrU3}31.5 & \cellcolor{lrU3}39.3 & \cellcolor{lrU5}52.4 & \cellcolor{lrU7}\textbf{78.6} \\
    \end{tabularx}
    \caption*{$\Frfdc = 4.9152$\,GS/s, \mycode{sc8} (16\,bit)}
  \end{subfigure}\\[0.6mm]
  \begin{subfigure}[t]{0.325\textwidth}
    \centering
    \begin{tabularx}{\linewidth}{@{}r@{\hspace{1.5pt}}|*{12}{>{\raggedleft\arraybackslash}X}@{}}
        $N$ & \rotatebox{90}{120} & \rotatebox{90}{200} & \rotatebox{90}{240} & \rotatebox{90}{300} & \rotatebox{90}{400} & \rotatebox{90}{480} & \rotatebox{90}{600} & \rotatebox{90}{800} & \rotatebox{90}{960} & \rotatebox{90}{1200} & \rotatebox{90}{1600} & \rotatebox{90}{2400} \\
        \hline
        1 & \cellcolor{lrU0}3.8 & \cellcolor{lrU0}6.4 & \cellcolor{lrU0}7.7 & \cellcolor{lrU0}9.6 & \cellcolor{lrU1}12.8 & \cellcolor{lrU1}15.4 & \cellcolor{lrU1}19.2 & \cellcolor{lrU2}25.6 & \cellcolor{lrU3}30.7 & \cellcolor{lrU3}38.4 & \cellcolor{lrU5}51.2 & \cellcolor{lrU7}\textbf{76.8} \\
        2 & \cellcolor{lrU0}7.7 & \cellcolor{lrU1}12.8 & \cellcolor{lrU1}15.4 & \cellcolor{lrU1}19.2 & \cellcolor{lrU2}25.6 & \cellcolor{lrU3}30.7 & \cellcolor{lrU3}38.4 & \cellcolor{lrU5}51.2 & \cellcolor{lrU6}61.4 & \cellcolor{lrU7}\textbf{76.8} & \cellcolor{lrOver}102.4 & \cellcolor{lrOver}153.6 \\
    \end{tabularx}
    \caption*{$\Frfdc = 4.8$\,GS/s, \mycode{sc16} (32\,bit)}
  \end{subfigure}\hfill
  \begin{subfigure}[t]{0.325\textwidth}
    \centering
    \begin{tabularx}{\linewidth}{@{}r@{\hspace{1.5pt}}|*{12}{>{\raggedleft\arraybackslash}X}@{}}
        $N$ & \rotatebox{90}{120} & \rotatebox{90}{200} & \rotatebox{90}{240} & \rotatebox{90}{300} & \rotatebox{90}{400} & \rotatebox{90}{480} & \rotatebox{90}{600} & \rotatebox{90}{800} & \rotatebox{90}{960} & \rotatebox{90}{1200} & \rotatebox{90}{1600} & \rotatebox{90}{2400} \\
        \hline
        1 & \cellcolor{lrU0}2.9 & \cellcolor{lrU0}4.8 & \cellcolor{lrU0}5.8 & \cellcolor{lrU0}7.2 & \cellcolor{lrU0}9.6 & \cellcolor{lrU1}11.5 & \cellcolor{lrU1}14.4 & \cellcolor{lrU1}19.2 & \cellcolor{lrU2}23 & \cellcolor{lrU2}28.8 & \cellcolor{lrU3}38.4 & \cellcolor{lrU5}\textbf{57.6} \\
        2 & \cellcolor{lrU0}5.8 & \cellcolor{lrU0}9.6 & \cellcolor{lrU1}11.5 & \cellcolor{lrU1}14.4 & \cellcolor{lrU1}19.2 & \cellcolor{lrU2}23 & \cellcolor{lrU2}28.8 & \cellcolor{lrU3}38.4 & \cellcolor{lrU4}46.1 & \cellcolor{lrU5}57.6 & \cellcolor{lrU7}\textbf{76.8} & \cellcolor{lrOver}115.2 \\
    \end{tabularx}
    \caption*{$\Frfdc = 4.8$\,GS/s, \mycode{sc12} (24\,bit)}
  \end{subfigure}\hfill
  \begin{subfigure}[t]{0.325\textwidth}
    \centering
    \begin{tabularx}{\linewidth}{@{}r@{\hspace{1.5pt}}|*{12}{>{\raggedleft\arraybackslash}X}@{}}
        $N$ & \rotatebox{90}{120} & \rotatebox{90}{200} & \rotatebox{90}{240} & \rotatebox{90}{300} & \rotatebox{90}{400} & \rotatebox{90}{480} & \rotatebox{90}{600} & \rotatebox{90}{800} & \rotatebox{90}{960} & \rotatebox{90}{1200} & \rotatebox{90}{1600} & \rotatebox{90}{2400} \\
        \hline
        1 & \cellcolor{lrU0}1.9 & \cellcolor{lrU0}3.2 & \cellcolor{lrU0}3.8 & \cellcolor{lrU0}4.8 & \cellcolor{lrU0}6.4 & \cellcolor{lrU0}7.7 & \cellcolor{lrU0}9.6 & \cellcolor{lrU1}12.8 & \cellcolor{lrU1}15.4 & \cellcolor{lrU1}19.2 & \cellcolor{lrU2}25.6 & \cellcolor{lrU3}\textbf{38.4} \\
        2 & \cellcolor{lrU0}3.8 & \cellcolor{lrU0}6.4 & \cellcolor{lrU0}7.7 & \cellcolor{lrU0}9.6 & \cellcolor{lrU1}12.8 & \cellcolor{lrU1}15.4 & \cellcolor{lrU1}19.2 & \cellcolor{lrU2}25.6 & \cellcolor{lrU3}30.7 & \cellcolor{lrU3}38.4 & \cellcolor{lrU5}51.2 & \cellcolor{lrU7}\textbf{76.8} \\
    \end{tabularx}
    \caption*{$\Frfdc = 4.8$\,GS/s, \mycode{sc8} (16\,bit)}
  \end{subfigure}\\[0.6mm]
  \vspace{-3.0mm}
  \caption{One-direction I/Q payload rate $\Rotw = N \cdot \Fsamp \cdot b$ (Gb/s) on the (a)~ZCU216 and (b)~RFSoC~4x2 for each RFDC rate $\Frfdc$ (rows of tables) and wire format (columns of tables).
  In each table, the rows are the channel count $N$ and the columns are the per-channel I/Q sample rates $\Fsamp$ ({\msps}) of Table~\ref{tab:rfsoc-board-config}.
  Cell colors show the utilization of the 100\,GbE link in 10\% steps, from light blue (0--10\%) to dark blue (90--100\%); red cells exceed the link, and the bold cell in each row is the highest rate that fits.}
  \label{fig:link_rates}
  \vspace{-2.0mm}
\end{figure*}

\subsection{Board Control Plane}
\label{sec:design:daemon}

A Python daemon on the RFSoC's Arm processor manages board discovery, ownership, and configuration.
It communicates with the host module over the 1\,GbE management network, separately from the high-speed 100\,GbE link carrying samples.
The daemon responds to device discovery and grants control to one client at a time, so competing applications cannot change DAC/ADC or FPGA settings while a device is in use.
For each UHD configuration request from the host, it programs the RFSoC's PLL clocks, DAC/ADC settings, fabric clocks, FPGA control registers, and performs multi-tile synchronization (MTS) when multiple channels are used.
A carrier frequency or I/Q sample rate change within the loaded bitstream retunes the RFDC's NCO or rate in place; while changing the RFDC rate or channel count requires loading another bitstream and reconfiguring the board, which takes less than ten seconds.

\subsection{Host: A DPDK UHD Device Module}
\label{sec:design:host}

The host side is a UHD device module that registers a new device type and exposes the RFSoC board through the standard \mycode{uhd} API.
Existing UHD applications can therefore interact with RFSoC boards without modifying UHD or their application code.
We design the same module to serve both x86 and Arm hosts, with architecture-specific kernels for sample conversion.
The host data path combines DPDK packet processing with an OTW converter that translates between interleaved I/Q samples and per-channel application buffers.

\myparatight{Packet processing.}
The host device module uses DPDK~\cite{dpdk} to access network interface card (NIC) queues directly from dedicated polling cores, bypassing the kernel network stack for data traffic while leaving management traffic on the ordinary network path.
Lock-free single-producer, single-consumer rings transfer packet-buffer ownership between polling cores and conversion workers without an additional payload copy at those handoffs.
This module also recycles the most recently freed buffers first, using a last-in, first-out (LIFO) policy to improve cache locality.

\myparatight{OTW converter and channel (de)interleaving on x86 and Arm hosts.}
Since the wire packets interleave samples from $N$ channels, whereas UHD applications exchange per-channel buffers, the OTW converter running on the host performs format conversion and (de-)interleaving in one pass, avoiding an intermediate buffer and an additional pass over the samples.
{\projectcode} supports this OTW converter and channel (de)interleaving with two vector implementations: AVX-512 on x86 and Neon on Arm (aarch64).
On x86 hosts, the module checks for AVX-512 at runtime and falls back to using scalar loops on a CPU without it; every aarch64 CPU has Neon, so Arm hosts always use the vector kernels.
For example, with the \mycode{sc16} wire format, a vector register holds 16 I/Q samples (16 pairs of 16-bit I and Q) with AVX-512 and 4 with Neon.
On RX, a kernel loads $N$ registers of consecutive I/Q samples from the packet, which hold the next 16 (or 4) samples of each of the $N$ channels, and de-interleaves them in $\log_2 N$ stages using vector shuffles and permutations.
Eventually, each register holds the samples of one channel, and the kernel converts and stores only the channels the application requested, writing them directly into application buffers.
On TX, the inverse stages interleave the channels directly into packet payloads; channels the application does not send are filled with zeros.
\mycode{sc8} samples are widened to 16 bits when loaded and, on TX, narrowed to 8 bits with saturation when stored.
sc12 samples are unpacked into sc16 samples on RX, or packed from them on TX.
The host kernels support $N = 1, 2, 4, 8, 16$: RX converts \mycode{sc16}, \mycode{sc12}, or \mycode{sc8} wire samples to \mycode{sc16} or \mycode{fc32} samples to be consumed by the application; TX performs the corresponding reverse conversions.
Thus, applications can retain their preferred sample representation while selecting a different wire precision to fit the host-device data link capacity.
The scalar loops produce the same bits and also convert the last samples of a call that do not fill a vector register, so both vector implementations are bit-identical to the scalar reference.

\myparatight{Thread scheduling.}
Streaming threads are pinned to reserved cores, with unrelated work and interrupts directed elsewhere to reduce potential scheduling conflicts.
On heterogeneous hosts, the host module ranks cores by their reported capacity and places the application thread, DPDK polling threads, and conversion workers on the higher-capacity cores. 
For example, on the DGX Spark with a GB10 superchip, these threads use the Cortex-X925 (performance) cores, leaving the Cortex-A725 (efficiency) cores available for other tasks.
Conversion workers also remain active while waiting for a timed transmit start, avoiding wake-up delays that could interrupt delivery to the FPGA's buffer in the TX streamer.

\section{Implementation and Hardware Platforms}
\label{sec:impl}

\begin{figure}[!t]
    \centering
    \begin{subfigure}[b]{0.48\columnwidth}
        \centering
        \includegraphics[width=\linewidth]{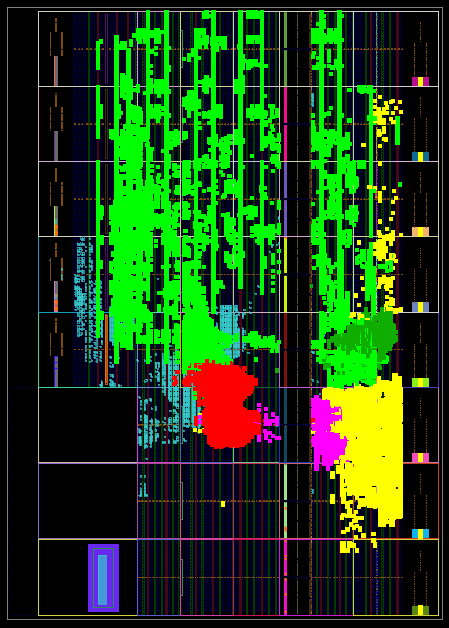}
        \caption{ZCU216, 16 channels.}
        \label{fig:fpga-layout:zcu216}
    \end{subfigure}\hspace{0.02\columnwidth}%
    \begin{subfigure}[b]{0.48\columnwidth}
        \centering
        \includegraphics[width=\linewidth]{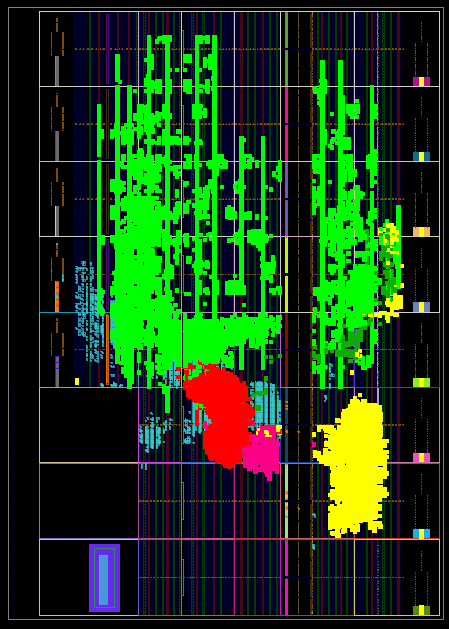}
        \caption{RFSoC~4x2, 2 channels.}
        \label{fig:fpga-layout:rfsoc4x2}
    \end{subfigure}
    \vspace{-3.0mm}
    \caption{FPGA designs of the two bitstreams in Table~\ref{tab:fpga-util}, including the TX streamer (\emph{light green}), TX phaser (\emph{dark green}), RX streamer (\emph{red}), RX phaser (\emph{pink}), RF DACs and ADCs (\emph{yellow}), and the rest of the circuits (\emph{light blue}).}
    \label{fig:fpga-layout}
    \vspace{-5.0mm}
\end{figure}

\myparatight{Implementation.}
{\projectcode} consists of the three components described in \S\ref{sec:design}, none of which modifies UHD.
The host module, \mycode{libuhd-rfsoc.so}, is implemented in about 8.8K lines of C++ code and built against UHD's public headers and DPDK~23.11.4; UHD loads it from its module path, where it registers the find and make functions for the \mycode{zcu216} and \mycode{rfsoc4x2} device types. 
The same source is also built for x86 and Arm hosts with AVX-512 and Neon conversion kernels.
The board daemon is about 9K lines of Python code on the PYNQ image of each board (2.7.0 on the ZCU216, 3.1 on the RFSoC~4x2) and drives the FPGA over AXI-Lite.
The FPGA data plane is a parameterized block design written in SystemVerilog, built with Vivado~2021.1 (ZCU216) or 2024.1 (RFSoC~4x2) into one bitstream for each (\mycode{board\_type}, $\Frfdc$, $N$) configuration, i.e., the 21 rows of Table~\ref{tab:rfsoc-board-config}, plus one fixed-rate two-channel RFSoC~4x2 bitstream at $\Frfdc = 5.0$\,GS/s and $\Fintdec = 2$ that streams 2.5\,GS/s per channel in \mycode{sc8} only. 
For example, the 16-channel ZCU216 design with $\Frfdc = 2.5\,\text{GS/s}$ (312.5\,{\msps} per channel) uses 7.0\% of the LUTs, 6.1\% of the flip-flops, and 3.7\% of the DSP slices of the device (Table~\ref{tab:fpga-util}).
Its on-chip memory usage, 51\% of the BRAM and 80\% of the URAM, comes mostly from the 4\,MiB TX buffer and is therefore the same for the 2-channel RFSoC~4x2 design.
Fig.~\ref{fig:fpga-layout} shows the actual FPGA resource layout including: TX streamer, TX phaser, RX streamer, RX phaser, RF DACs and ADCs, and the rest of the circuits on both devices.

\begin{table}[!t]
    \centering
    \caption{FPGA resource utilization of two example bitstreams: the 16-channel ZCU216 design with $\Frfdc = 2.5\,\text{GS/s}$ and the 2-channel RFSoC~4x2 design with $\Frfdc = 5.0\,\text{GS/s}$.}
    \label{tab:fpga-util}
    \vspace{-3.0mm}
    \footnotesize
    \setlength{\tabcolsep}{10pt}
    \renewcommand{\arraystretch}{1.15}
    \begin{tabular}{@{}lrrr@{}}
        \toprule
        \textbf{Resource} & \textbf{Available} & \textbf{ZCU216} & \textbf{RFSoC~4x2} \\
        \midrule
        LUT    & 425,280   & 29,705 (6.98\%)   & 28,279 (6.65\%) \\
        LUTRAM & 213,600   & 1,050 (0.49\%)    & 795 (0.37\%) \\
        FF     & 850,560   & 51,838 (6.09\%)   & 45,898 (5.40\%) \\
        BRAM   & 1,080     & 551.5 (51.06\%)   & 551.5 (51.06\%) \\
        URAM   & 80        & 64 (80.00\%)      & 64 (80.00\%) \\
        DSP    & 4,272     & 158 (3.70\%)      & 72 (1.69\%) \\
        GT     & 16        & 4 (25.00\%)       & 4 (25.00\%) \\
        BUFG   & 696       & 10 (1.44\%)       & 9 (1.29\%) \\
        MMCM   & 8         & 1 (12.50\%)       & 1 (12.50\%) \\
        \bottomrule
    \end{tabular}
    \vspace{-3.0mm}
\end{table}

\myparatight{Hardware platforms.}
We consider two RFSoC boards (ZCU216 and RFSoC~4x2) and two host types (x86 and Arm).
The ZCU216 (XCZU49DR with 16 ADCs and 16 DACs) is connected to the host through a QSFP28-to-4xSFP28 breakout cable, and the RFSoC~4x2 (XCZU48DR with 4 ADCs and 2 DACs) is connected to the host through its QSFP28 port, both at 100\,GbE, with a separate 1\,GbE management network carrying control traffic.
The x86 host is a 6-core, 12-thread Xeon w3-2423 CPU with a single 32\,GB DDR5 RDIMM at 4400\,MT/s (single memory channel with a peak bandwidth of 35.2\,GB/s), an NVIDIA ConnectX-5 dual-port 100\,GbE adapter on PCIe~3.0 x16, and runs Ubuntu~24.04 (Linux 5.15) with UHD~4.6.0.0 and DPDK~23.11.4, and GNU Radio~3.10.9.2.
The Arm host is an NVIDIA DGX Spark with a GB10 superchip with 10 Cortex-X925 (high-performance) and 10 Cortex-A725 (high-efficiency) cores, 128\,GB of unified memory (shared between CPU and GPU), Ubuntu~24.04 OS, and a ConnectX-7 adapter supporting dual-port 200\,GbE.
Since the RFSoC~4x2 board supports up to only 100\,GbE, we set the DGX Spark's NIC port to that rate.

\section{Evaluation}
\label{sec:eval}

We evaluate {\projectcode} at four levels, covering I/Q sample transport, application compatibility, end-to-end communication DSP pipelines, and full-stack RAN operation.
At the \emph{UHD level}, we measure the host OTW converter's throughput in GS/s per core and evaluate full-rate, full-duplex streaming reliability using UHD's \mycode{benchmark\_rate}.
At the \emph{GNU~Radio level}, we demonstrate I/Q sample streaming through unmodified \mycode{gr-uhd} blocks~\cite{gnuradio_gruhd} and validate runtime per-channel phase control from the received time-domain waveforms and spectra.
At the \emph{end-to-end communication DSP level}, we evaluate a customized, GPU-based Zak-OTFS transceiver~\cite{zheng2026realtimescalablezakotfsreceiver} running on top of {\projectcode} and report its communication performance.
At the \emph{full-stack RAN level}, we run the OCUDU~\cite{ocudu} and Duranta~\cite{duranta} gNBs on a host with an RFSoC board as its radio through {\projectcode}, and demonstrate communication with both srsUE-based software UEs and commodity UEs.
The software UEs likewise can run on hosts connected to RFSoC radios through {\projectcode}.
Across these four levels, the experiments cover different combinations of RFSoC boards as well as both x86 and Arm hosts.

\subsection{Microbenchmarks}

\begin{figure}[!t]
    \centering
    \includegraphics[width=0.98\columnwidth]{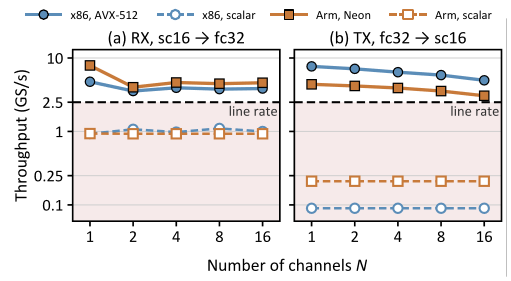}
    \vspace{-3.0mm}
    \caption{Throughput of {\projectcode}'s host OTW converter with varying channel count $N$ for (a) \texttt{sc16} to \texttt{fc32} for the RX, and (b) \texttt{fc32} to \texttt{sc16} for the TX, on one core of the Xeon w3-2423 (x86, AVX-512 kernels) and on one Cortex-X925 performance core of the GB10 (Arm, Neon kernels), each against its scalar loops.
    Each point covers the format conversion, the (de-)interleaving of the $N$ channels, and the copy between the packet payload and the per-channel buffers, in GS/s of I/Q samples across all channels.
    The dashed line is the aggregated $\Frfdc = {2.5}\,\text{GS/s}$ line rate per direction.}
    \label{fig:kernels}
    \vspace{-5.0mm}
\end{figure}

\myparatight{Host OTW converter throughput.}
Fig.~\ref{fig:kernels} shows the benchmarked throughput of the host's OTW converter between the \mycode{sc16} wire format and \mycode{fc32}, the floating-point format commonly used by UHD applications, across varying numbers of channels, $N$, in both directions.
The benchmark uses a payload of 64 packets of 2,048 \mycode{sc16} samples (512\,KB) and counts the I/Q samples of all $N$ channels converted per second using a single thread on the host CPU, averaged over 3 seconds of repeated conversion.
With AVX-512, a single core converts {3.55--4.75}\,{\gsps} on RX and {4.97--7.69}\,{\gsps} on TX, which is {1.4--3.1}$\times$ the required aggregated sample rate at $\Frfdc=2.5\,\text{GS/s}$ per direction.
The scalar implementation reaches only {0.93--1.10}\,{\gsps} on RX and 0.09\,{\gsps} on TX, where it reproduces the rounding rule bit-identical to the vector kernels.
Overall, AVX-512 is {3.3--5.1}$\times$ faster on RX and about {55--85}$\times$ faster on TX.
On one Cortex-X925 performance core of the GB10 in DGX Spark, Neon converts {4.00--7.93}\,{\gsps} on RX and {3.06--4.39}\,{\gsps} on TX, {1.2--3.2}$\times$ the required rate and {4.3--8.5}$\times$ and {15--21}$\times$ faster than its scalar loops, with output bit-identical to the scalar reference.
Overall, the conversion rate varies moderately with $N$.

\myparatight{Full-duplex streaming robustness.}
We test with \mycode{benchmark\_rate}, the stock streaming benchmark from the official UHD repository~\cite{ettus_uhd}, which streams TX and RX samples at a given rate and counts every overflow (i.e., RX samples lost because the host fell behind), underflow (i.e., the TX buffer ran dry), dropped sample, sequence error, and late packet. 
A testing run passes successfully only if it reports \emph{zero} for every error type during 30 seconds of full-rate TX and RX streaming across all channels. 
For each of the 21 board configurations of Table~\ref{tab:rfsoc-board-config} and each wire format (\mycode{sc16}, \mycode{sc12}, and \mycode{sc8}), we test the highest per-channel I/Q sample rate that fits the 100\,GbE host-device link.
This yields 63 test cases: 21 board configurations (15 on the ZCU216 and 6 on the RFSoC~4x2), each in three wire formats.
Ten runs are performed for each test case on both the x86 and Arm hosts, yielding $63 \times 10 \times 2 = 1{,}260$ runs, all with zero errors.

\subsection{GNU Radio Flowgraphs}
\label{sec:eval:gnuradio}

\begin{figure*}[!t]
    \centering
    \begin{subfigure}[b]{0.215\textwidth}
        \centering
        \includegraphics[width=\linewidth]{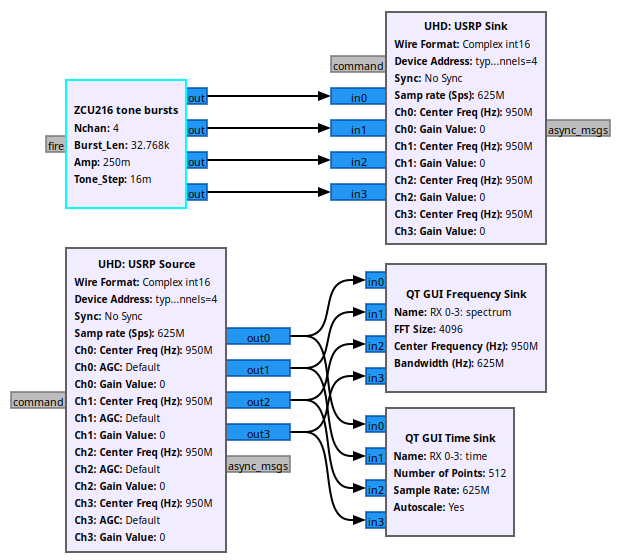}
        \caption{Flowgraph.}
        \label{fig:gnuradio:gui}
    \end{subfigure}\hfill
    \begin{subfigure}[b]{0.385\textwidth}
        \centering
        \includegraphics[width=\linewidth]{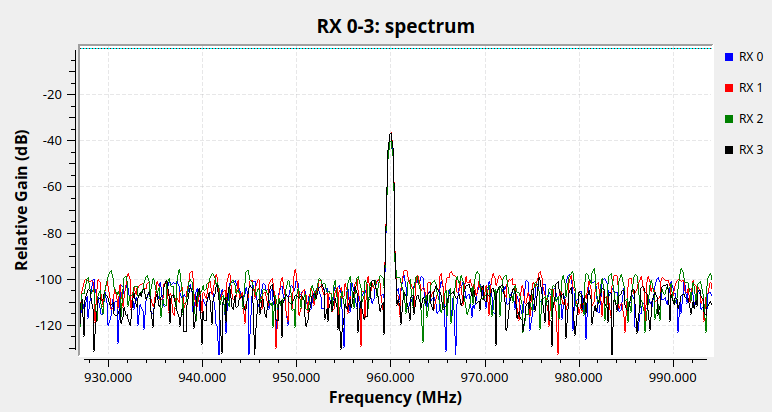}
        \caption{Spectrum of RX 0--3.}
        \label{fig:gnuradio:freq}
    \end{subfigure}\hfill
    \begin{subfigure}[b]{0.385\textwidth}
        \centering
        \includegraphics[width=\linewidth]{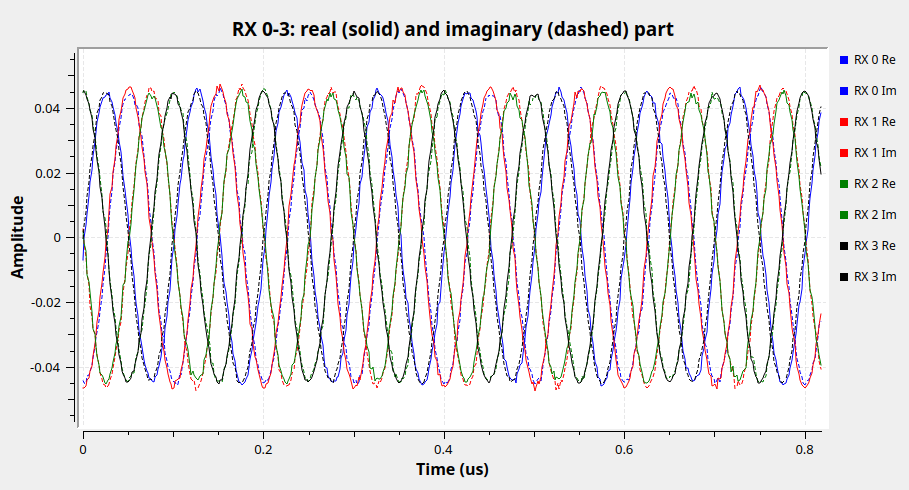}
        \caption{RX 0--3 with the TX phases tuned 90$^\circ$ apart.}
        \label{fig:gnuradio:phase}
    \end{subfigure}
    \vspace{-3.0mm}
    \caption{The 4-channel GNU Radio example on the ZCU216, run on the x86 host through {\projectcode} at 625\,{\msps} per channel in the \texttt{sc16} wire format, with a 10\,{MHz} tone at {950}\,{MHz} carrier frequency.}
    \label{fig:gnuradio}
    \vspace{-3.0mm}
\end{figure*}

We provide GNU~Radio Companion flowgraphs for both boards across their supported channel counts.
Fig.~\ref{fig:gnuradio}(a) shows the 4-channel (4TX, 4RX) ZCU216 example, running at 625\,{\msps} I/Q sample rate per channel with the \mycode{sc16} wire format on an x86 host connected over 100\,GbE.
Each DAC output is connected by an RF cable to the corresponding ADC input, forming four loopback paths.
The flowgraph uses \emph{standard} UHD USRP Sink and Source blocks configured with \mycode{type=zcu216}, QT GUI Sink blocks, and an embedded Python block that generates a 10\,MHz complex-valued sinusoid.
With $\Fcarr = {950}\,\text{MHz}$ carrier setting, the transmitted RF tone is at 960\,MHz, and the 930--990\,MHz window shown in Fig.~\ref{fig:gnuradio}(b) lies well inside the first Nyquist zone.
To accommodate GUI processing at this sample rate, the flowgraph operates in timed bursts.
Every 0.5\,s, it schedules a 32,768-sample TX burst per channel and performs RX capture for each channel within the transmit interval.
Each capture contains contiguous samples at the full configured sample rate, while the gaps between bursts reduce the average GNU~Radio and GUI processing load.
A common RX start timestamp selects the same acquisition window on all synchronized channels.

Fig.~\ref{fig:gnuradio}(b) shows the received spectra
around 960\,MHz, with the tone peaks more than 60\,dB
above the displayed noise floor.
The GUI also provides two tabs with per-channel sliders for configuring the TX phase (from $-180^\circ$ to $180^\circ$) and amplitude (from 0 to 1 at full scale), using the per-channel phaser feature (\S\ref{sec:design:fpga}).
Users can adjust these controls while the flowgraph runs and observe the changes in the received time-domain waveforms.
Fig.~\ref{fig:gnuradio}(c) shows four received tones with approximately equal amplitudes and a $90^\circ$ phase difference between successive channels.

\subsection{GPU-based Zak-OTFS Transmission}
\label{sec:eval:otfs}

Orthogonal time frequency space (OTFS) modulation~\cite{mohammed2022otfs} places information symbols on a two-dimensional delay-Doppler (DD) grid, offering robustness to time-varying multipath channels.
Zak-OTFS maps this DD grid directly to time-domain waveforms using the inverse discrete Zak transform.
However, channel estimation and equalization become computationally demanding with increased DD grid sizes, motivating GPU-based receiver processing that exploits the structured sparsity of the DD-domain channel~\cite{zheng2026realtimescalablezakotfsreceiver}.
We use this workload to evaluate {\projectcode} with an end-to-end, GPU-accelerated DSP pipeline.

We run the Zak-OTFS transceiver on an  NVIDIA DGX Spark connected to an RFSoC~4x2 board through {\projectcode}.
The RFSoC~4x2's TX and RX ports are connected by an RF cable to form a SISO loopback link at $\Fcarr = 1\,\text{GHz}$ carrier frequency, with an I/Q sample rate of 122.88\,{\msps}, a DD grid with dimension (4096,32), and QPSK/16QAM modulation.
Each packet contains a Zadoff-Chu (ZC) preamble followed by one pilot frame and one data frame.
The transmitter generates the two frames by applying the inverse discrete Zak transform to a DD-domain point pilot and a grid of modulation symbols, respectively.
The receiver uses the ZC preamble for per-packet timing synchronization and carrier frequency offset (CFO) estimation and correction.
The GPU then performs discrete Zak transforms, pilot-based DD channel estimation, MMSE equalization using a structured-sparse conjugate-gradient solver, and symbol demodulation.
Our implementation uses FP16 processing and CUDA graph replay to reduce processing and kernel launch overhead.

The results show that the link achieves throughput of {115}\,{Mb/s} with QPSK and {230}\,{Mb/s} with 16QAM, while the DGX Spark processes the received packets, streamed through {\projectcode} in real time.
We also sweep the link SNR and measure BER over 300 packets per point.
For QPSK, we observe no bit errors at the tested SNRs of 30\,dB and above; the first nonzero BER is approximately $10^{-7}$ at 27\,dB, increasing to $10^{-2}$ at 17\,dB.
For 16QAM, we observe no bit errors at the tested SNRs of 40\,dB and above; the first nonzero BER is approximately $10^{-6}$ at 34\,dB, increasing to $10^{-3}$ at 30\,dB.
At the maximum transmit level, with an estimated receive SNR of 52\,dB, the measured EVM is 2.0\% for both
modulations.
These results show a custom GPU-based PHY integrated with an RFSoC radio through {\projectcode}.

\subsection{5G RAN Stacks: OCUDU and Duranta}
\label{sec:eval:ocudu}

\begin{figure*}[!t]
    \centering
    \includegraphics[width=0.99\textwidth]{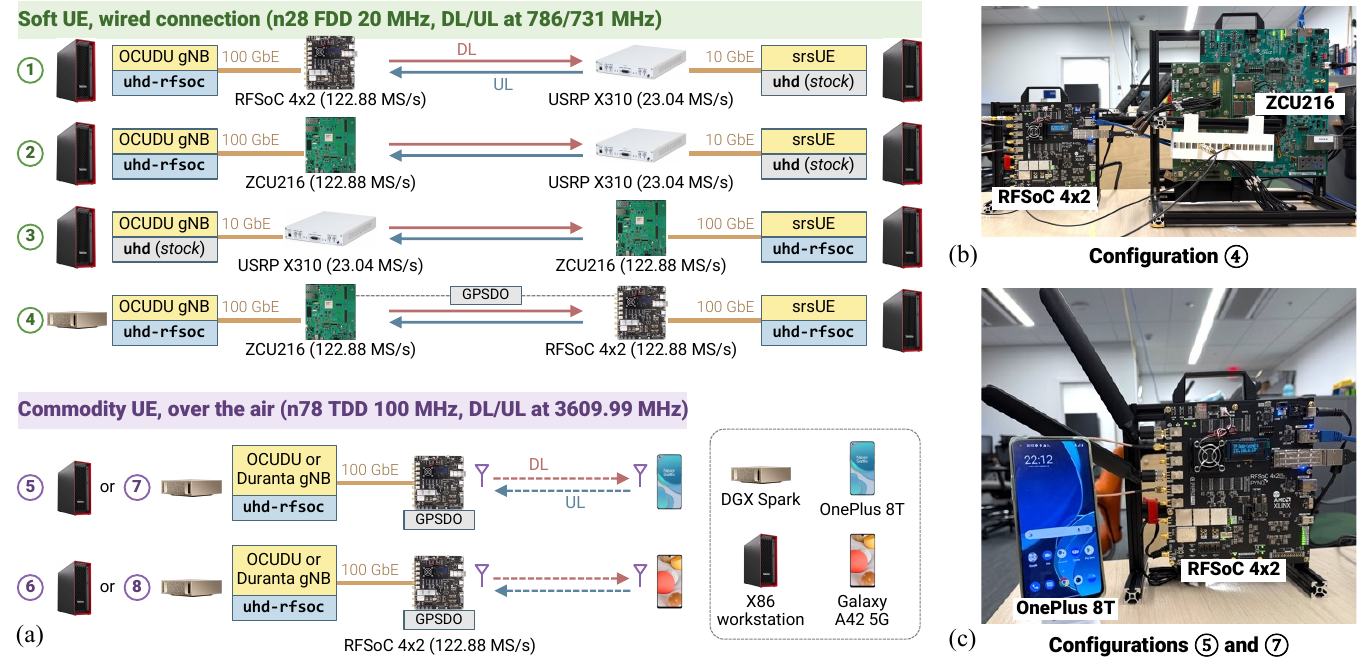}
    \vspace{-3.0mm}
    \caption{(a) Tested gNB--UE configurations for the OCUDU and Duranta 5G RAN stacks.
    (b) The RFSoC~4x2 and ZCU216 boards connected over a cable.
    (c) The over-the-air setup using the RFSoC~4x2 as the gNB and the OnePlus 8T.}
    \label{fig:ocudu_setup}
    \vspace{-3.0mm}
\end{figure*}

At the RAN software level, we run two open-source 5G RAN stacks, OCUDU~\cite{ocudu} and Duranta~\cite{duranta}, with the RFSoC boards as their radios through {\projectcode}.
The OCUDU or Duranta gNB runs unmodified in Split~8, whose radio is the ZCU216 or RFSoC~4x2 board through {\projectcodetitle}, or a USRP X310 through UHD.
The gNB host is an x86 workstation (Lenovo P5) running AVX-512 kernels, or the Arm-based NVIDIA DGX Spark running Neon kernels.
Fig.~\ref{fig:ocudu_setup}(a) shows the eight gNB--UE configurations \circledwhite{1}--\circledwhite{8}, and Figs.~\ref{fig:ocudu_setup}(b) and~(c) show the setups of \circledwhite{4} and \circledwhite{5}--\circledwhite{8}, respectively.

\myparatight{Wired connections (Configurations 1--4).}
The UE is srsUE~\cite{srsran4g} with the X310 or an RFSoC board as its radio, and the transmitter of each radio is connected to the receiver of the other using SMA cables.
In \circledwhite{1}--\circledwhite{3}, the gNB and srsUE run on the same x86 workstation, each with its own radio and network interface; in \circledwhite{4}, the gNB runs on the DGX Spark and srsUE on the x86 workstation.
The cell is a 20\,MHz FDD cell in NR band n28 (15\,kHz subcarrier spacing, DL 786\,MHz, UL 731\,MHz).
\circledwhite{1}--\circledwhite{3} combine each RFSoC board with the X310, as either the gNB or the UE radio; while \circledwhite{4} connects the two RFSoC boards to each other, the ZCU216 as the gNB radio and the RFSoC~4x2 as the UE radio, both locked to a shared GPS-disciplined 10\,MHz reference (GPSDO).

The two radios connected by a wire run at different I/Q sample rates in Fig.~\ref{fig:ocudu_setup}(a): the X310 at 23.04\,{\msps}, and the RFSoC boards at 122.88\,{\msps}, the lowest LTE/5G~NR rate that both boards provide (see Table~\ref{tab:rfsoc-board-config}).
The gNB and UE software stacks absorb the mismatch in different ways, and neither needs a change in {\projectcodetitle}, which only serves the rate that each stack requests through the standard UHD call.
Specifically, OCUDU runs its lower PHY at the sample rate of its radio: the rate must be a multiple of 1.92\,MHz, sample-accurate for the physical random access channel (PRACH) cyclic prefix, and at least the channel bandwidth, and the OFDM modulator and demodulator use a DFT of the rate divided by the subcarrier spacing, 8192 points at 122.88\,{\msps} and 15\,kHz, where the 106 resource blocks of the 20\,MHz cell occupy 1272 subcarriers (4,096 points for the 100\,MHz cell at 30\,kHz of the over-the-air pairs).
srsUE instead keeps its NR PHY at a fixed rate and resamples in its radio layer by an integer factor with an FFT-based resampler: with the X310, the PHY and the radio both run at 23.04\,{\msps} (1,536-point DFT); with an RFSoC board as the UE radio (\circledwhite{3} and \circledwhite{4}), the radio runs at 122.88\,{\msps} and the PHY at 30.72\,{\msps}, which requires setting the NR PHY rate in srsUE independently of the RFSoC radio's I/Q sample rate.

\myparatight{Over-the-air connection (Configurations~5--8).}
In this case, the UE is a commodity phone, a OnePlus~8T in \circledwhite{5} and \circledwhite{7}, or a Samsung Galaxy~A42~5G in \circledwhite{6} and \circledwhite{8}, and the gNB radio is the RFSoC~4x2 at 122.88\,{\msps}, its clock locked to the 10\,MHz output of a GPS-disciplined oscillator (GPSDO).
The cell is a 100\,MHz TDD cell in band n78 (30\,kHz subcarrier spacing) at 3609.99\,MHz, in the second Nyquist zone of the RFSoC~4x2.
The gNB host is the x86 workstation in \circledwhite{5} and \circledwhite{6} and the DGX Spark in \circledwhite{7} and \circledwhite{8}.

\begin{table}[!t]
    \centering
    \caption{End-to-end throughput of the gNB--UE configurations in Fig.~\ref{fig:ocudu_setup}, measured with iperf3 (for the phones: UDP at a target bit rate of 200\,Mb/s offered in each direction).
    DL and UL throughput are measured in (Mb/s), and RTT measures the ping round-trip time in ms.}
    \label{tab:ocudu}
    \vspace{-3.0mm}
    \footnotesize
    \setlength{\tabcolsep}{3pt}
    \renewcommand{\arraystretch}{1.15}
    \begin{tabular}{@{}rlllcc@{}}
        \toprule
        \textbf{\#} & \textbf{Host} & \textbf{gNB} & \textbf{UE} & \textbf{DL/UL (Mb/s)} & \textbf{RTT (ms)} \\
        \midrule
        \multicolumn{6}{@{}l}{\textbf{OCUDU gNB + soft UE} (over a cable)} \\
         1 & x86   & RFSoC~4x2  & X310           & 46.0/74.4   & 31.8 \\
         2 & x86   & ZCU216     & X310           & 57.9/71.7   & 31.3 \\
         3 & x86   & X310       & ZCU216         & 56.0/74.0   & 31.2 \\
         4 & DGX~Spark & ZCU216     & RFSoC~4x2      & 57.9/73.9   & 30.9 \\
        \midrule
        \multicolumn{6}{@{}l}{\textbf{OCUDU gNB + commodity UE} (over the air)} \\
         5 & x86   & RFSoC~4x2  & OnePlus 8T     & 168.1/166.7 & 31.2 \\
         6 & x86   & RFSoC~4x2  & Galaxy A42 5G  & 191.3/152.0 & 29.0 \\
         7 & DGX~Spark & RFSoC~4x2  & OnePlus 8T     & 200.0/162.8 & 31.5 \\
         8 & DGX~Spark & RFSoC~4x2  & Galaxy A42 5G  & 196.6/140.8 & 29.8 \\
        \midrule
        \multicolumn{6}{@{}l}{\textbf{Duranta gNB + commodity UE} (over the air)} \\
         5 & x86   & RFSoC~4x2  & OnePlus 8T     & 163.4/92.7  & 21.8 \\
         6 & x86   & RFSoC~4x2  & Galaxy A42 5G  & 200.0/132.7 & 20.3 \\
         7 & DGX~Spark & RFSoC~4x2  & OnePlus 8T     & 179.5/109.1 & 20.6 \\
         8 & DGX~Spark & RFSoC~4x2  & Galaxy A42 5G  & 200.0/119.8 & 20.3 \\
        \bottomrule
    \end{tabular}
    \vspace{-3.0mm}
\end{table}

\myparatight{Results with OCUDU.}
Table~\ref{tab:ocudu} shows the throughput and round-trip time of each pair.
For wired connections, \circledwhite{2} and \circledwhite{3} show that the ZCU216 serves equally well as the gNB radio and as the UE radio: the downlink reaches 57.9\,{Mb/s} and 56.0\,{Mb/s} and the uplink 71.7\,{Mb/s} and 74.0\,{Mb/s}, close to the capacity of the 20\,{MHz} cell, with a round-trip time (RTT) of about 31\,{ms}.
\circledwhite{1} with the RFSoC~4x2 as the gNB radio, reaches 46.0\,{Mb/s} in the downlink at modulation and coding scheme (MCS) 22--26 and 74.4\,{Mb/s} in the uplink from the X310 at a physical uplink shared channel (PUSCH) SNR of 37\,dB. 
When the gNB runs on the DGX Spark in \circledwhite{4}, it reaches 57.9\,{Mb/s} in the downlink and 73.9\,{Mb/s} in the uplink, both at MCS~28, with an RTT of about 31\,{ms}.
Over the air, we measure the phones using iperf3 over UDP with a target bit rate of 200\,{Mb/s} in each direction for 20\,s, and report the rate at the receiver.
The phones in \circledwhite{5}--\circledwhite{8} reach 168.1--200.0\,{Mb/s} in the downlink and 140.8--166.7\,{Mb/s} in the uplink through the 100 MHz cell, with an RTT of 29--32\,{ms}.
With the DGX Spark as the gNB host, the Arm host and its Neon kernels can sustain the 122.88\,{\msps} gNB radio as the x86 host does.

\myparatight{Results with Duranta.}
Table~\ref{tab:ocudu} also lists the throughput and RTT of the over-the-air configurations.
Specifically, the phones reach 163.4--200.0\,{Mb/s} in the downlink and 92.7--132.7\,{Mb/s} in the uplink, with RTT of 20--22\,{ms}.
The x86 workstation delivers 163.4\,{Mb/s} and 200.0\,{Mb/s} in the downlink and 92.7\,{Mb/s} and 132.7\,{Mb/s} in the uplink (\circledwhite{5} and \circledwhite{6}).
The DGX Spark delivers 179.5\,{Mb/s} and 200.0\,{Mb/s} in the downlink and 109.1\,{Mb/s} and 119.8\,{Mb/s} in the uplink (\circledwhite{7} and \circledwhite{8}).
Compared to OCUDU under the same configurations, Duranta has a 9--11\,{ms} shorter RTT, while OCUDU reaches the higher uplink rate in all four configurations; in the downlink, the Galaxy~A42~5G reaches 191.3--200.0\,{Mb/s} and the OnePlus~8T 163.4--200.0\,{Mb/s} across the two stacks.
Both stacks drive the same RFSoC~4x2 through {\projectcode}, so these differences come from the stacks' schedulers and configurations rather than from the radio.

\section{Conclusions}
\label{sec:conclusion}

We presented {\projectcode}, an open-source FPGA design and host software stack that brings off-the-shelf RFSoC boards, the ZCU216 and RFSoC~4x2, into the UHD ecosystem without modifying UHD.
Through standard UHD calls and APIs, applications can configure the channel count, carrier frequency, per-channel I/Q sample rate, and the wire format. 
Over one 100\,GbE host-device link, {\projectcode} streams 2.5\,{\gsps} of 16-bit or 5.0\,{\gsps} of 8-bit I/Q samples in each direction across up to 16 channels.
All supported board configurations pass UHD's built-in streaming benchmark in full duplex on both x86 and Arm hosts. 
Existing UHD-based software also runs on the boards unmodified, including
GNU~Radio flowgraphs built from stock blocks, a GPU-based Zak-OTFS transceiver DSP pipeline, as well as the open-source OCUDU and Duranta 5G RAN stacks.
These results show that off-the-shelf RFSoC boards can be integrated into the existing wireless ecosystem through the UHD interface.

\begin{acks}
This work was supported in part by NSF under grants 2211944, 2434131, 2443137, 2450567, 2549359, and 2554552, and by ARO under Awards No.~W911NF-26-2-A158 and W911NF-26-2-A164.
This work was also supported in part by the Center for Ubiquitous Connectivity (CUbiC), sponsored by Semiconductor Research Corporation (SRC) and Defense Advanced Research Projects Agency (DARPA) under the JUMP 2.0 program.
We thank Neel Pandeya (National Instruments) and Ivan Seskar (Rutgers University) for their helpful discussions and constructive feedback.
\end{acks}

\bibliographystyle{ACM-Reference-Format}
\bibliography{references}

\end{document}